\documentclass[pdftex,twocolumn,epjc3]{svjour3}          

\usepackage{amssymb}
\usepackage{amsmath}
\usepackage{subcaption}
\usepackage{snapshot}

\RequirePackage[T1]{fontenc} 

\smartqed  

\RequirePackage{graphicx}
\RequirePackage{mathptmx}      
\RequirePackage{flushend}
\RequirePackage[numbers,sort&compress]{natbib}
\RequirePackage[colorlinks,citecolor=blue,urlcolor=blue,linkcolor=blue]{hyperref}

\journalname{Nikhef 2019-043}

\begin{document}

\title{\boldmath Triple Higgs boson production to six $b$-jets at a 100 TeV proton collider}

\subtitle{Anomalous self-couplings and gauge-singlet scalars}

\author{Andreas Papaefstathiou\thanksref{e1,addr1, addr2}
        \and
        Gilberto Tetlalmatzi-Xolocotzi\thanksref{e2,addr2}
        \and
        Marco Zaro\thanksref{e3,addr2}
}

\thankstext{e1}{e-mail: apapaefs@cer
n.ch}
\thankstext{e2}{e-mail: gtx@nikhef.nl}
\thankstext{e3}{e-mail: m.zaro@nikhef.nl}

\institute{Institute for Theoretical Physics Amsterdam and Delta
  Institute for Theoretical Physics, University of Amsterdam, Science
  Park 904, 1098 XH Amsterdam, The Netherlands\label{addr1}
  \and
  Nikhef, Theory Group, Science Park 105, 1098 XG, Amsterdam, The
  Netherlands\label{addr2}
}

\date{\today}

\maketitle

\begin{abstract}
We investigate the production of three Higgs bosons at a proton-proton collider running at a centre-of-mass energy of 100~TeV, all of which decay into $b$-jets. This final state encapsulates by far the largest fraction of the total cross section of triple Higgs boson production, approximately $20\%$. We examine, by constructing detailed phenomenological analyses, two scenarios: (i) one in which the triple and quartic Higgs boson self-couplings are modified independently by new phenomena with respect to their Standard Model (SM) values and (ii) an extension of the SM by a gauge-singlet scalar that could drive first-order electroweak phase transition, within the context of the so-called xSM. In the former, we find that competitive constraints of $\mathcal{O}(1)$ can be placed on the quartic coupling and in the latter we demonstrate that it will be possible to obtain important information on the structure of the extended scalar sector. 
\end{abstract}

\section{Introduction}
In the past decade of operation of CERN's Large Hadron Collider (LHC), the landscape of particle physics has changed dramatically. The discovery of the Higgs boson and the lack of stark signals of new phenomena around the TeV scale are defining characteristics of this new era. In the years to come the Higgs boson is set to become itself a tool for exploration and discovery. This will be particularly true at the future circular collider (FCC), which is planned to be hosted in a 100 km tunnel, envisioning an ensemble of $e^+e^-$, $e^+p$ and $pp$ collider programmes through towards the end of the 21st century~\cite{Contino:2016spe, Mangano:2016jyj, Abada:2019zxq, Benedikt:2018csr, Abada:2019lih}. Taken together, all of these programmes aim to map the properties of the Higgs boson and the electroweak gauge bosons with an accuracy order(s) of magnitude better than today and to improve by almost an order of magnitude the discovery reach for new particles. 

A particular ``flagship'' target of the FCC will be the investigation of the Higgs potential, through the measurement of the Higgs boson's ($h$) self-interactions that can be written, post-electroweak symmetry breaking (EWSB), as:
\begin{equation}\label{eq:hpotential}
V(h) = \frac{1}{2} m_h^2 h^2 + \lambda_3 v_0 h^3 + \frac{1}{4} \lambda_4 h^4 \,,
\end{equation}
\noindent where $v_0 \simeq 246$~GeV is the Higgs vacuum expectation value (vev), $m_h \simeq 125$~GeV is the Higgs boson mass and the self-couplings take the values $\lambda_3 =  \lambda_4 =  m_h^2 / 2v_0^2 \equiv \lambda_\mathrm{SM}$ within the SM. Legacy LHC measurements are expected to provide an $\lesssim \mathcal{O}(1)$ measurement of the triple coupling, $\lambda_3$, with respect to its SM value~\cite{ATL-PHYS-PUB-2018-053,Cepeda:2019klc}, and no significant direct information on the quartic self-coupling $\lambda_4$. On the other hand, several studies have demonstrated the potential of the proton-proton programme of the FCC (the FCC-hh), to constrain the triple coupling to within a few percent of the SM value, particularly through the production of Higgs boson pairs~\cite{Azatov:2015oxa, Contino:2016spe, Papaefstathiou:2015iba, Lu:2015jza, He:2015spf, Cao:2016zob, Banerjee:2018yxy, Chang:2018uwu}. Several studies have also hinted that constraints are possible on the quartic coupling at the FCC-hh, either indirectly in double Higgs boson production~\cite{Bizon:2018syu, Borowka:2018pxx}, or directly through triple Higgs boson production~\cite{Plehn:2005nk, Papaefstathiou:2015paa, Chen:2015gva, Fuks:2017zkg, Dicus:2016rpf, Agrawal:2017cbs, Kilian:2017nio}. Up until now, in the case of the latter process, the following final states have been considered:
\begin{itemize}
\item $ hhh \rightarrow (b\bar{b}) (b\bar{b}) (\gamma\gamma)$,
\item  $hhh \rightarrow (b\bar{b}) (b\bar{b})(\tau^+\tau^-)$, 
\item $ hhh \rightarrow (b\bar{b}) (\tau^+\tau^-) (\tau^+\tau^-)$,
\item$hhh \rightarrow  (b\bar{b})$ $(W^+W^+) (W^+W^-)$. 
\end{itemize}
The sum of all these channels represents less than 10\% of the total branching ratio of $hhh$. In the present article, we investigate for the first time, to the best of our knowledge, the process that encapsulates by far the largest branching ratio: the case in which all three Higgs bosons decay into bottom quarks ($b\bar{b}$), resulting in complex final states involving six $b$-jets.\footnote{We note that the equivalent final state in Higgs boson pair pair production, leading to 4 $b$-jets, has been considered extensively in both phenomenological and experimental studies, see e.g.~\cite{deLima:2014dta, Wardrope:2014kya, Behr:2015oqq, Aad:2015uka, Khachatryan:2015yea, Aaboud:2016xco, Aaboud:2018knk, Li:2019tfd, Alves:2019igs}.}


In addition to understanding EWSB, non-standard Higgs boson self-couplings might provide the first experimental evidence of extra gauge-singlet scalars at the weak scale. These new scalar particles could ``catalyse'' electroweak phase transition, turning it into a violent, out-of-equilibrium event accompanied by massive entropy production (a first-order transition), enabling electroweak baryogenesis and thus explaining the observed matter-antimatter asymmetry, see e.g.~\cite{Espinosa:1993bs, Espinosa:2007qk, Barger:2007im, Espinosa:2008kw, Espinosa:2011ax, Cline:2012hg}. Evidence of such phenomena in multi-scalar production processes could materialise, for example, even in the case where the mixing of this new scalar and the ``SM-like'' Higgs boson is small. Indeed, current limits put an upper bound to the mixing angle of $\cos \theta \gtrsim 0.85$ and at the end of the high-luminosity run of the LHC this is expected to be $\gtrsim 0.95$~\cite{Profumo:2014opa}. First indications of the existence of these singlets could arise in resonant SM-like Higgs boson pair production for example, either at later stages of the LHC or during the FCC-hh lifetime. Such signals, along with the measurement of the SM-like Higgs self-coupling through non-resonant Higgs boson pair production, may not be sufficient to understand the nature of the additional singlet scalar. The production of three of these scalar particles, such as triple SM-like Higgs boson production, the main object of this article, could provide essential additional information both on the triple scalar couplings and on the quartic couplings. We demonstrate that this is possible by employing the six $b$-jet final state that maximises the cross section.\footnote{We would like to note here that the six $b$-jet final state might be also interesting in the context of $hh+Z$, see e.g.~\cite{Nordstrom:2018ceg}, or any triple neutral boson final state.}

The article is organised as follows: in section~\ref{sec:pheno} we discuss the setup used and describe the phenomenological analysis in the context of triple SM Higgs boson production. In section~\ref{sec:d4c3} we discuss the constraints that can be obtained in the anomalous coupling picture, where the self-couplings are rescaled with respect to the SM values, and in section~\ref{sec:singlet} we investigate in explicit benchmark scenarios, the potential for discovering triple Higgs boson production in the presence of a singlet scalar that can viably generate first-order electroweak phase transition, taken from~\cite{Kotwal:2016tex}. We conclude in section~\ref{sec:conclusions}. In~\ref{app:variations} we provide investigations of relevant uncertainties entering our analysis.

\section{Searching for triple Higgs boson production}\label{sec:pheno}

\subsection{The setup}
In what follows, we generate parton-level events either at leading order or next-to-leading order by using \texttt{MadGraph5} \texttt{\_aMC@NLO} \cite{Alwall:2014hca, Hirschi:2015iia} and shower/match them via the MC@NLO method~\cite{Frixione:2002ik} where appropriate, via the \texttt{HERWIG} (7.1.5) parton shower~\cite{Bahr:2008pv, Gieseke:2011na, Arnold:2012fq, Bellm:2013hwb, Bellm:2015jjp, Bellm:2017bvx}. We include modeling of the hadronization and the underlying event but no detector effects beyond geometry. We use the parton density function set \texttt{NNPDF23\_lo} \texttt{\_as\_0130\_qed}~\cite{Ball:2012cx} throughout the chain of event generation. 

For the analysis, we cluster final-state particles with transverse momentum $p_T > 100$~MeV into anti-$k_T$ jets~\cite{Cacciari:2008gp} with radius parameter $R=0.4$ via the \texttt{FastJet} package~\cite{Cacciari:2011ma}. We use the \texttt{HwSim} package~\cite{hwsim} for \texttt{HERWIG} to write out event files for each sample in a custom compressed \texttt{ROOT} format~\cite{Brun:1997pa} and to perform the phenomenological analysis. 

\subsection{\boldmath Differential distributions in $pp \rightarrow hhh$ at 100 TeV}

In the present subsection we investigate the form of differential distributions for the $hhh$ signal within the SM. Variations of the shapes of these distributions due to the effect of new phenomena are considered in the respective sections below: in section~\ref{sec:d4c3} we show variations due to different values of the anomalous couplings and in section~\ref{sec:singlet} we show variations in the presence of a singlet scalar at various masses. We refer the reader to~\cite{Papaefstathiou:2015paa} for additional distributions, including a comparison to Higgs boson pair production at 100~TeV. 

\begin{figure}
  \includegraphics[width=0.9\columnwidth]{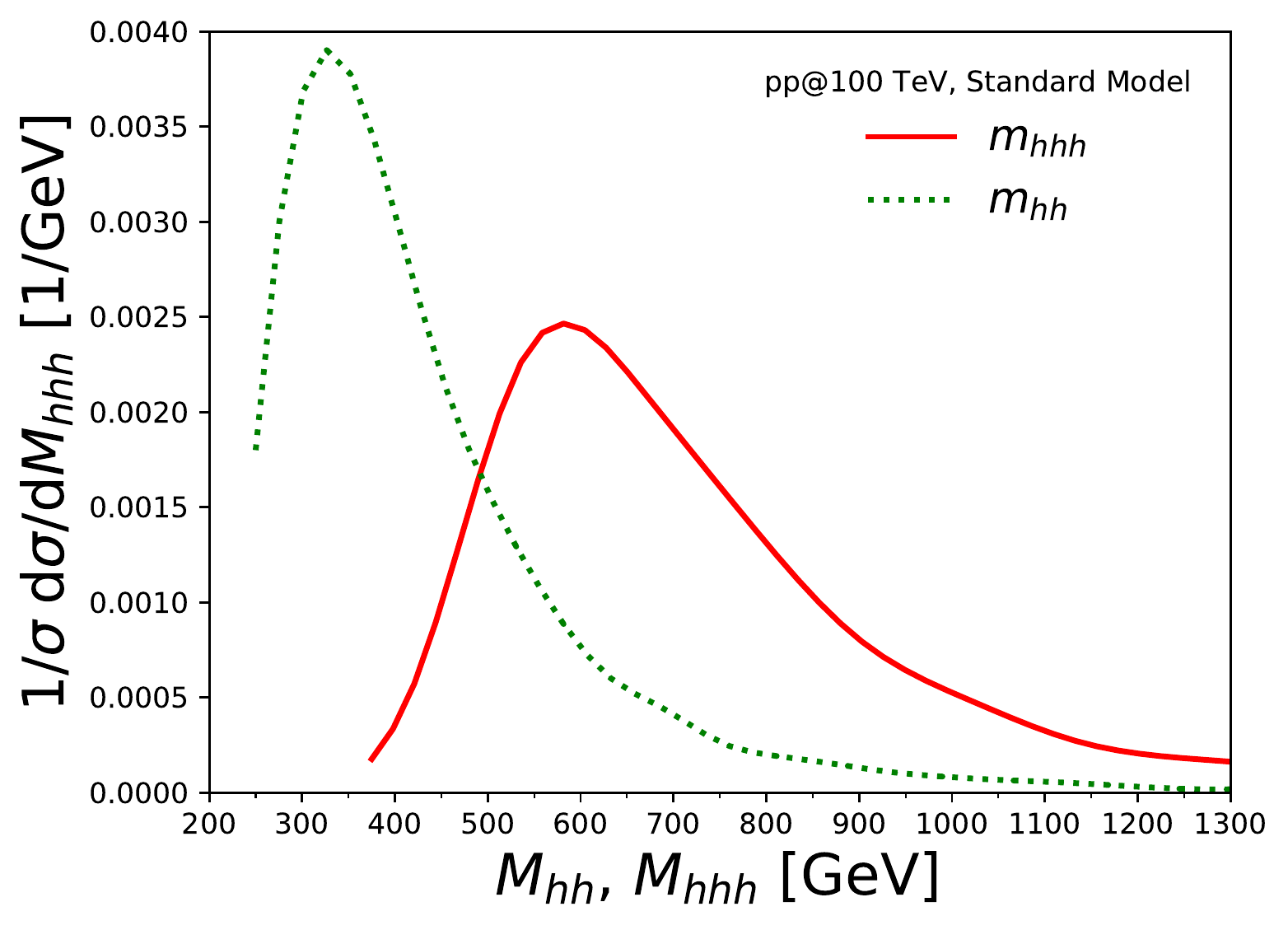}
\caption{The invariant mass of \textit{any} two ($hh$, dotted green) or all three ($hhh$, solid red) Higgs bosons in triple Higgs production at the FCC-hh at 100 TeV, reconstructed from Monte Carlo truth with no cuts applied. }
\label{fig:invmass}
\end{figure}

\begin{figure}
  \includegraphics[width=0.9\columnwidth]{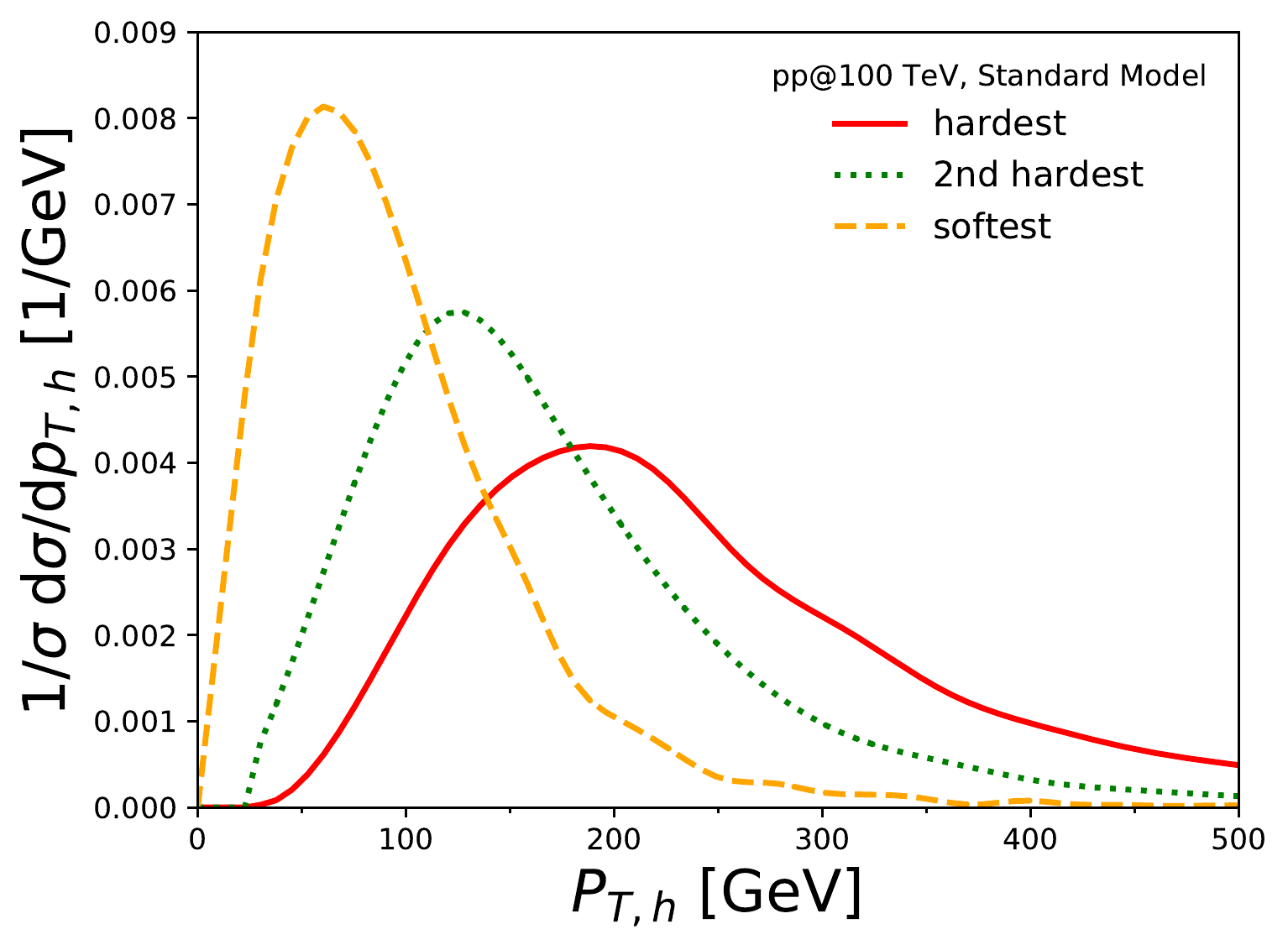}
\caption{The transverse momentum of Higgs bosons ordered from the hardest to softest (solid red, dotted green, dashed yellow, respectively), in triple Higgs production at the FCC-hh at 100 TeV, reconstructed from Monte Carlo truth with no cuts applied.}
\label{fig:pthiggs}
\end{figure}

We show in fig.~\ref{fig:invmass}, the invariant mass of (any) two or all three Higgs bosons reconstructed from Monte Carlo truth with no cuts applied, $M_{hh}$ and $M_{hhh}$, respectively. The former peaks at $\sim 300$~GeV whereas the latter at $\sim 600$~GeV. In fig.~\ref{fig:pthiggs} we show the Monte Carlo truth transverse momentum of the Higgs bosons ordered from hardest to softest. The transverse momentum distributions peak at $\sim 200$~GeV, $\sim 150$~GeV and $\sim 50$~GeV from hardest to softest, respectively. 

\subsection{Event generation}
The simulation of final states containing up to six coloured objects remains a challenge to this day, even at tree level. In the present study we provide initial estimates by considering the efficiency of a phenomenological analysis on $ (b\bar{b}) (b\bar{b}) (b\bar{b}) $ final states. We stress here that we have simulated the QCD-induced $ (b\bar{b}) (b\bar{b}) (b\bar{b}) $ exactly at tree level.\footnote{In particular, this was made possible thanks to the technique of Ref.~\cite{Hirschi:2015iia} that performs a Monte Carlo over helicities.} For backgrounds which arise from charm-jets or light jets being mis-identified as $b$-jets (i.e. the reducible backgrounds), we have estimated the cross sections and assumed the analysis efficiencies to be identical to the equivalent process with $b$-quarks, factoring out the mis-identification rates.  

For the irreducible backgrounds, i.e. those that constitute the ``real'' $ (b\bar{b}) (b\bar{b}) (b\bar{b}) $ final states, we have considered processes that contain three bosons (either a Higgs boson or a $Z$ boson) that each then decay into $(b\bar{b})$: $hhZ$, $hZZ$, $ZZZ$. We have included the loop-induced gluon-fusion component in the case of $hZZ$ and $ZZZ$.\footnote{All processes incorporate the full effect of spin correlations, implemented via \texttt{MadSpin}~\cite{Artoisenet:2012st} that follows the method of~\cite{Frixione:2007zp}, apart from $gg\rightarrow hZZ$ and $gg\rightarrow ZZZ$. This is not expected to have a significant impact on the analysis efficiencies.} Furthermore, we have considered backgrounds with either one or two bosons plus $(b\bar{b})$ that originate from QCD interactions: $hZ+(b\bar{b})$, $hh+(b\bar{b})$, $ZZ+(b\bar{b})$ and $Z+(b\bar{b})(b\bar{b})$, $h+(b\bar{b})(b\bar{b})$. Of the aforementioned processes, $Z+(b\bar{b})(b\bar{b})$ and $h+(b\bar{b})(b\bar{b})$ turn out to be the largest contributors to total background cross section. For the latter process, $h+(b\bar{b})(b\bar{b})$, we also consider the Higgs effective theory contributions (i.e. including the effective interaction $ggh$), which constitute approximately 
3/4 of the cross section.\footnote{A similar ratio was also observed for $h+(b\bar{b})$ at the LHC~\cite{Deutschmann:2018avk}.} However, we have found that the largest background component by far is the pure QCD production of $ (b\bar{b}) (b\bar{b}) (b\bar{b}) $. The details of the analysis are presented in the next subsection.

\begin{table}
\caption{The generation-level cuts imposed on the processes. The index $j$ indicates any quark flavour or gluons.}
\label{tab:gencuts}
\centering
\begin{tabular}{ll}
observable & cut \\
\noalign{\smallskip}\hline\noalign{\smallskip}
$p_{T,j}$& $>30~\mathrm{GeV}$\\
 $|y _j|$ & $< 5.0$\\ 
$\Delta R_{j,j}$ & $>0.2$  \\
\noalign{\smallskip}\hline
\end{tabular}
\end{table}

We have simulated the triple Higgs boson signal at (loop-induced) leading order and the quark-anti-quark-initiated component of the tri-boson processes at next-to-leading order. We have generated samples of $10^4$ events for all signal processes, except for the SM $hhh$, for which we generated $10^5$ events to obtain statistically reliable estimates of the significance. 

All other processes have been simulated at leading order. To take into account the higher-order corrections, we multiply all leading-order cross sections by a $K$-factor of 2. The size of the higher-order corrections is well-motivated for the $hhh$ signal by approximate calculations, see~\cite{Maltoni:2014eza}. In~\ref{app:variations}, we provide variations of the $K$-factor for the backgrounds to take into account this uncertainty, 
while given that a full NLO computation would be needed, we do not consider effects due to shapes. We have imposed generation-level cuts on processes that involve quarks of QCD origin. We list these cuts in table~\ref{tab:gencuts}. 

We emphasise that the simulation of processes with more than three final-state legs at next-to-leading order is an essential aspect that should be addressed in future studies at higher-energy hadron colliders, as such final states will become increasingly common. 

\subsection{Analysis details}\label{sec:anal}

We give here the details of the phenomenological hadron-level analysis that are common between the different new physics scenarios that we consider.

We ask for the events to contain exactly six identified $b$-jets with transverse momentum $p_T > 45$~GeV. We ask for these jets to lie within a pseudo-rapidity of $|\eta|  < 3.2$ and we also ask for the distance between any two $b$-jets to satisfy $\Delta R > 0.3$. The latter choice is simply to bring all processes on equal footing, given that the backgrounds that contain QCD-initiated $b$-quarks also obey a generation-level cut of $\Delta R > 0.2$. We consider the potential impact of reducing the pseudo-rapidity coverage for the identified $b$-jets on our conclusions in~\ref{app:variations}. For each of the 15 possible arrangements $I = \{ ij, kl, mn \}$ of the six $b$-jets into pairs we construct the observable:
\begin{equation}
\chi^2 = \sum_{qr \in \mathrm{pairings~} I} (M_{qr} - m_h^2)^2 \;,
\end{equation}
where $M_{qr}$ is the invariant mass of the $b$-jet pairing $qr$ in the arrangement of pairings $I$ and $m_h$ is the Higgs boson mass. Given that it is challenging to determine experimentally the charge of the $b$-quarks that initiated the $b$-jets, we consider the minimisation of the $\chi^2$ observable over all the possible pairings. The arrangement of pairings $I$ that gives the minimum of $\chi^2$, which we call $\chi^2_{\mathrm{min}}$, defines the three ``reconstructed Higgs bosons'', $h_r^i$, for $i=\{1,2,3\}$. For this specific combination we calculate the absolute difference with the Higgs mass and order from smallest to larger: ($\Delta m_\mathrm{min}$, $\Delta m_\mathrm{mid}$, $\Delta m_\mathrm{max}$). We impose cuts on the observables $\sqrt{\chi^2_{\mathrm{min}}}$, $\Delta m_\mathrm{min}$, $\Delta m_\mathrm{mid}$ and $\Delta m_\mathrm{max}$. Furthermore, we impose cuts on the transverse momentum of the hardest, second hardest and softest reconstructed Higgs boson, $p_T(h_r^i)$ for $i=\{1,2,3\}$. We also impose cuts on the distances between the reconstructed Higgs bosons, $\Delta R(h_r^i, h_r^j)$. Finally, we ask for the distances between the two $b$-jets that comprise the reconstructed Higgs bosons, $\Delta R_{bb}(h^i)$, to satisfy certain upper bounds. The values of the cuts on these observables are summarised in table~\ref{tab:analysiscuts}.

\begin{table}
\caption{The cuts that comprise the phenomenological analysis at hadron level.}
\label{tab:analysiscuts}
\centering
\begin{tabular}{ll}
observable & cut \\
\noalign{\smallskip}\hline\noalign{\smallskip}
$p_{T,b}$& $>45~\mathrm{GeV}$\\
 $|\eta _b|$ & $< 3.2$\\ 
$\Delta R_{b,b}$ & $>0.3$  \\
$p_T(h^i)$ & $> [170, 120, 0]$~GeV, $i=1,2,3$\\
$\chi^2_{\mathrm{min}}$ & $<17$~GeV \\
$\Delta m_\mathrm{min,~mid,~max}$ & $< 8, 8, 11$~GeV\\
$\Delta R(h_r^i, h_r^j) $ & $< [3.5, 3.5, 3.5]$,  $(i,j) = [ (1,2), (1,3), (2,3)]$ \\
$\Delta R_{bb}(h^i)$ & $< [3.5, 3.5, 3.5]$, $i=1,2,3$\\
\noalign{\smallskip}\hline
\end{tabular}
\end{table}

\section{Standard Model-like triple Higgs boson production}\label{sec:d4c3}

\subsection{Anomalous self-couplings}

\begin{figure*}[!htp]
 \centering
 \subcaptionbox{$\mathcal{M} \sim 1$\label{diag1}}{\includegraphics[width=0.22\textwidth]{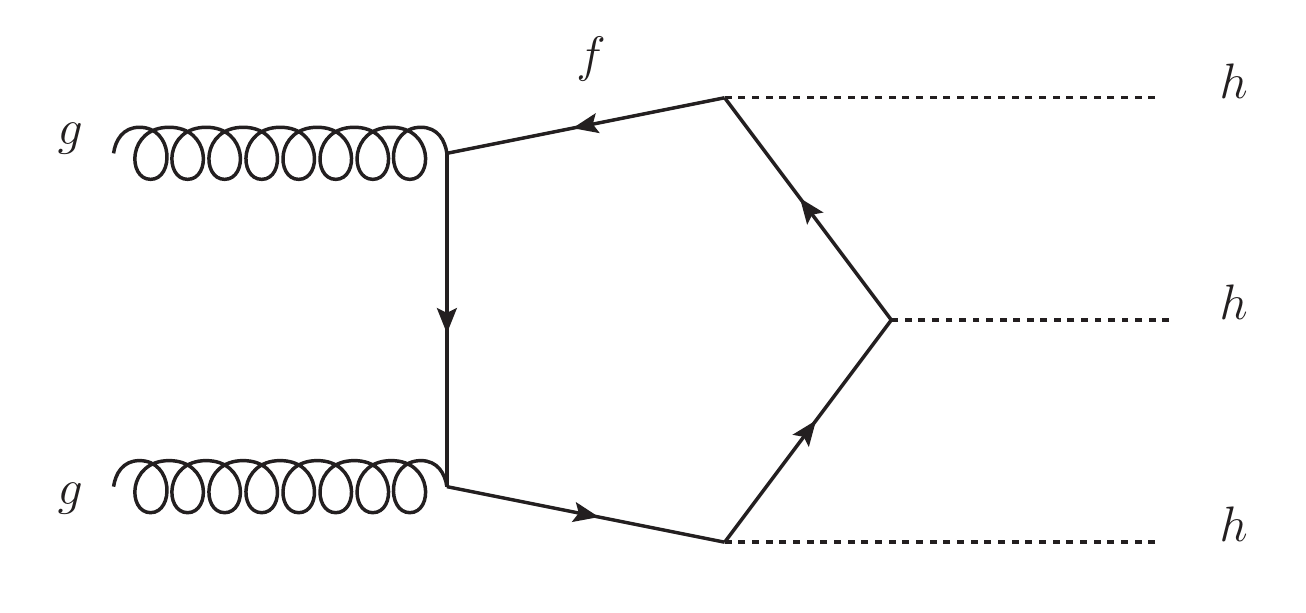}}
 \subcaptionbox{$\mathcal{M} \sim (1+c_3)$\label{diag2}}{\includegraphics[width=0.22\textwidth]{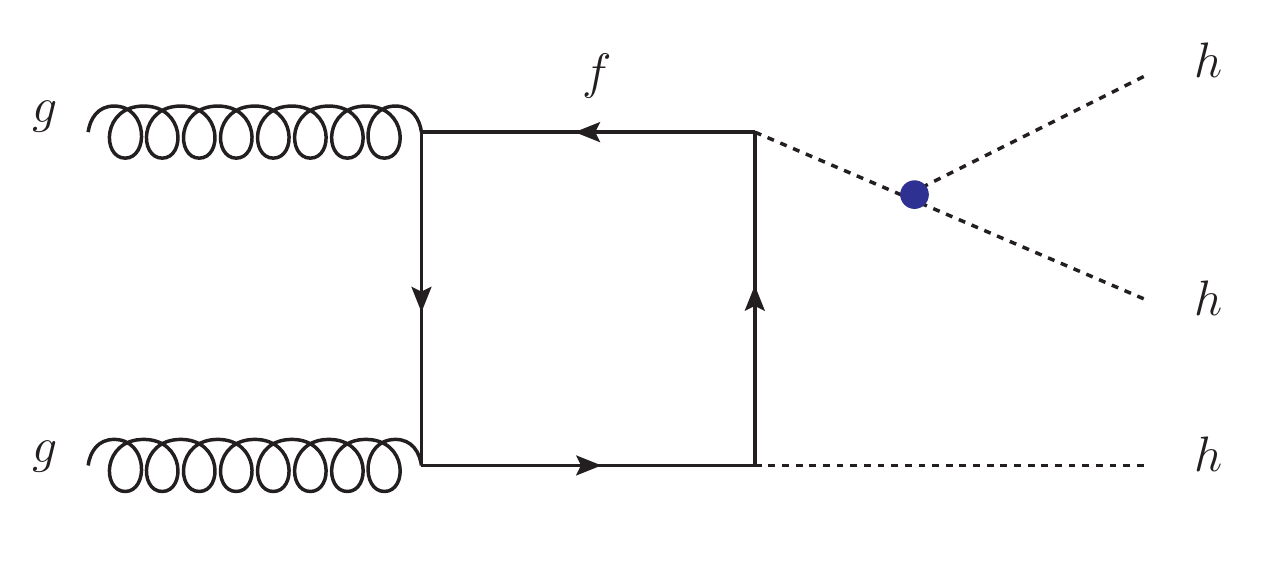}}
 \subcaptionbox{$\mathcal{M} \sim (1+d_4)$\label{diag3}}{\includegraphics[width=0.22\textwidth]{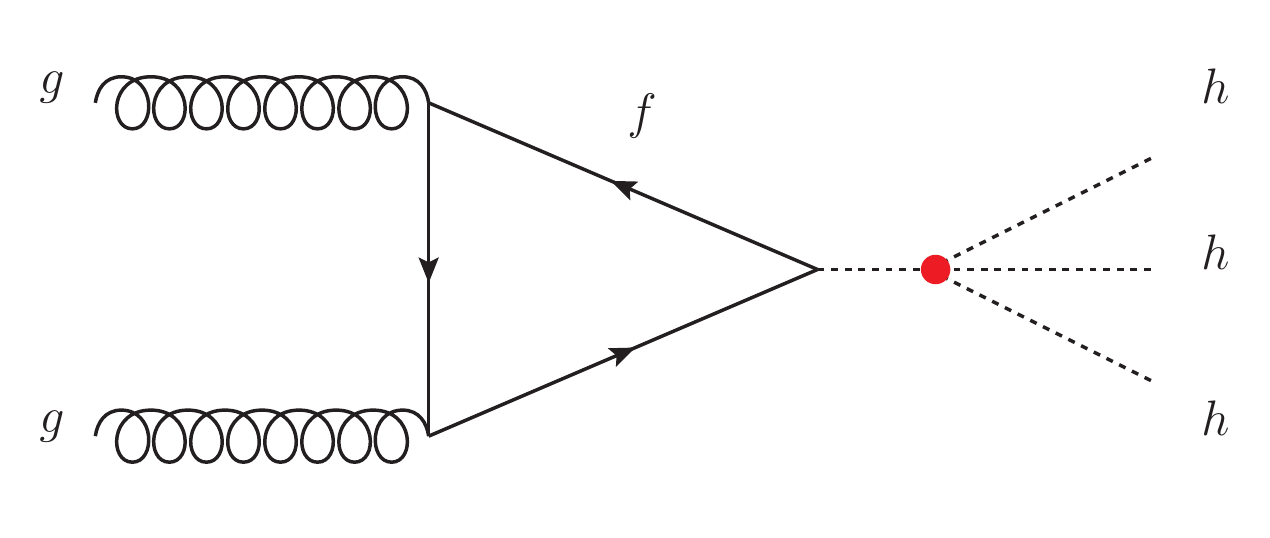}}
 \subcaptionbox{$\mathcal{M} \sim (1+c_3)^2$\label{diag4}}{\includegraphics[width=0.22\textwidth]{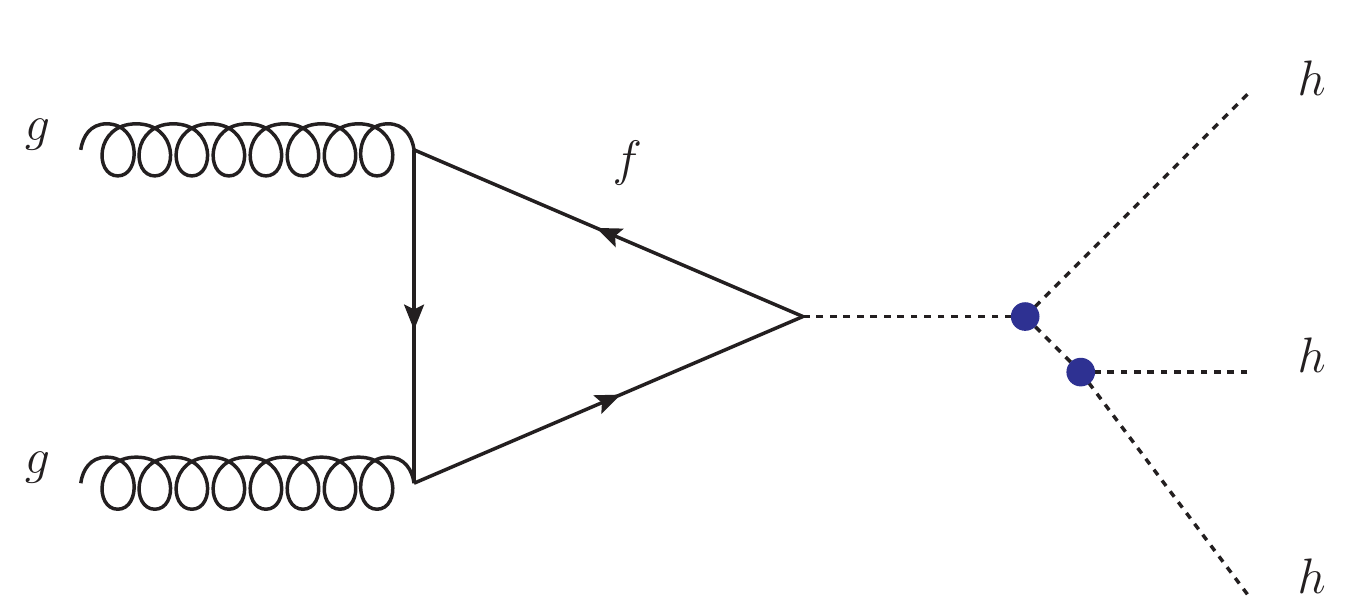}}
\caption{\label{fig:feyndiags} Example Feynman diagrams contributing to Higgs boson triple production via gluon fusion in the Standard Model, taken from~\cite{Papaefstathiou:2015paa}. The vertices highlighted with blobs indicate either triple (blue) or quartic (red) self-coupling contributions. In the $(c_3, d_4)$ model, diagram (a) produces matrix elements unmodified by the anomalous couplings (at this order) (b) produce terms $\propto (1+c_3)$, diagram (c) $\propto (1+d_4)$ and diagram (d) $\propto (1+c_3)^2$. The interference of such diagrams produces the terms that appear in eq.~\ref{eq:sigmac3d4}. }
\end{figure*}

We first consider a scenario in which the triple and quartic couplings are modified independently of each other. This ``agnostic'' anomalous coupling approach does not necessarily represent a physically viable theory, but allows for an investigation of the possible constraints that can be obtained for SM-like triple Higgs boson production. We thus consider interactions of the form: 
\begin{equation}\label{eq:c3d4}
V(h) = \frac{1}{2} m_h^2 h^2 + \lambda_{\mathrm{SM}} (1+c_3) v_0 h^3 + \frac{1}{4} \lambda_{\mathrm{SM}} ( 1 + d_4 ) h^4 \,,
\end{equation}
where the coefficients $c_3$ and $d_4$ represent the modifications of the triple and quartic Higgs boson self-interactions respectively. Assuming that the Yukawa couplings to the top and bottom quarks remain unchanged, these interactions will induce changes to the main production channel for triple Higgs boson production, that proceeds through gluon fusion, mediated by heavy quark loops. Example Feynman diagrams are shown in fig.~\ref{fig:feyndiags}, together with their scaling with the coefficients $c_3$ and $d_4$. 

\begin{figure}[t!]
  \includegraphics[width=\columnwidth]{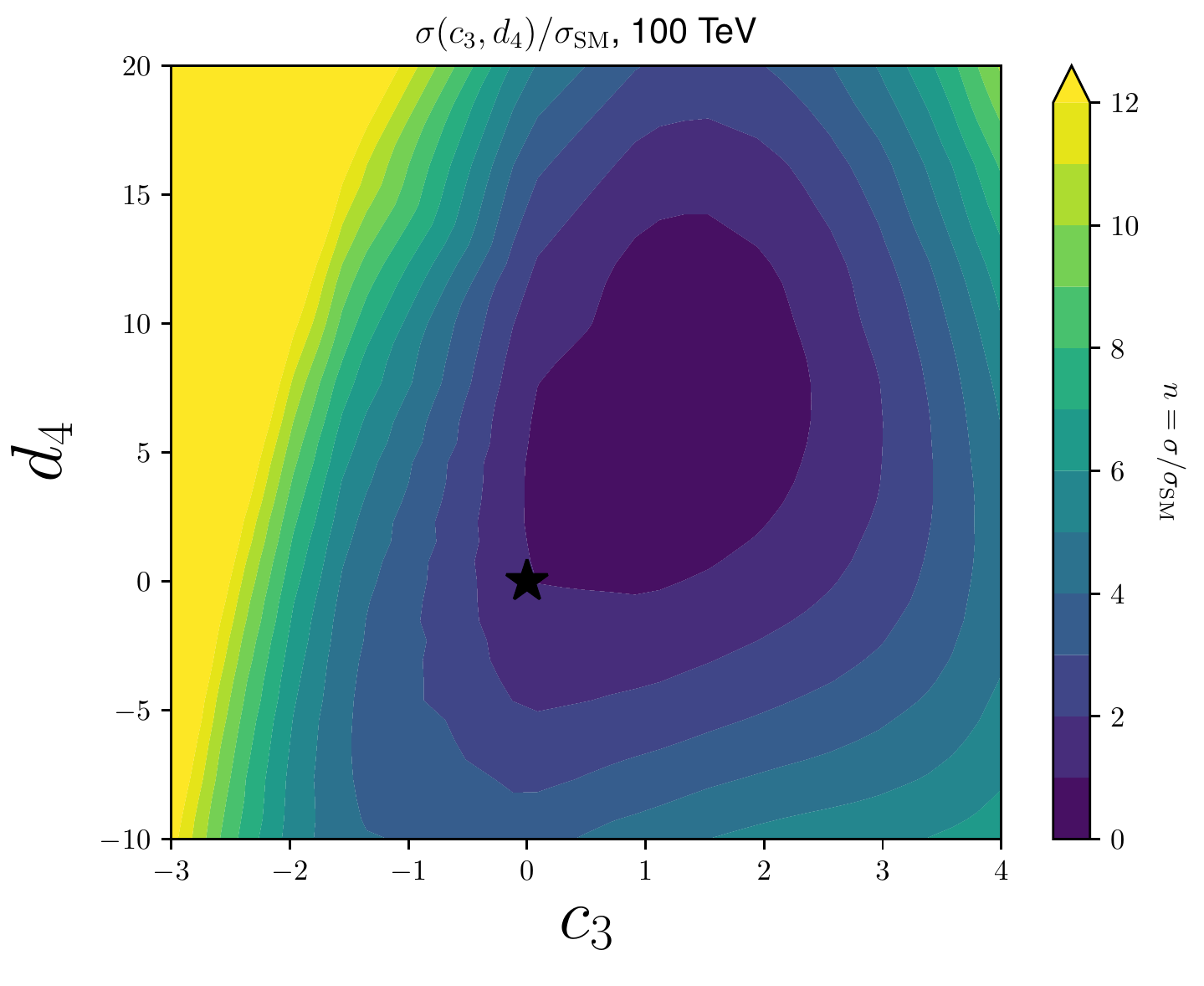}
\caption{The cross section for triple Higgs production with modified self-couplings ($\lambda_4 = \lambda_{\mathrm{SM}} (1+d_4)$ and $\lambda_3 = \lambda_{\mathrm{SM}} (1+c_3)$) at the FCC-hh at 100 TeV, normalised to the SM value. The black star indicates the SM point.}
\label{fig:xsc3d4}
\end{figure}

In fig.~\ref{fig:xsc3d4}  we show a variation of the cross section at a 100 TeV proton collider, normalised to the SM value. Evidently, variations of the triple self-coupling via $c_3$ produce larger changes than equivalent variations with $d_4$. A fit of the cross section on this plane yields a polynomial in $c_3$ and $d_4$ which is quartic in $c_3$ and quadratic in $d_4$. This is because there exist diagrams with two insertions of the triple self-coupling $c_3$ in triple Higgs boson production (diagram~\ref{diag4}), whereas there are only diagrams with at most a single insertion of $d_4$ (diagram~\ref{diag3}) at this order. The dependence of the cross section on $c_3$ and $d_4$, normalised to the SM cross section, was fitted as:
\begin{eqnarray}\label{eq:sigmac3d4}
\frac{\sigma(c_3, d_4)_{hhh}}{\sigma(\mathrm{SM})_{hhh}}-1  &=&  0.0309 \times c_3^4 - 0.2079 \times c_3^3 \nonumber \\
&+&  0.0407 \times c_3^2 d_4  + 0.7384\times c_3^2\nonumber\\
&+& 0.0156 \times d_4^2 - 0.1450 \times c_3 d_4\nonumber\\
&-& 0.1078 \times d_4 - 0.6887 \times c_3  \;.
\end{eqnarray}

The formula above can be used to estimate the cross section in any model with SM-like Higgs boson triple production. For example, in the context of the SM effective field theory at $D=6$, setting the relation $d_4 = 6 c_3$, one reproduces to a good approximation, the fit of~\cite{Papaefstathiou:2015paa}.

\begin{figure}[t!]
  \includegraphics[width=0.9\columnwidth]{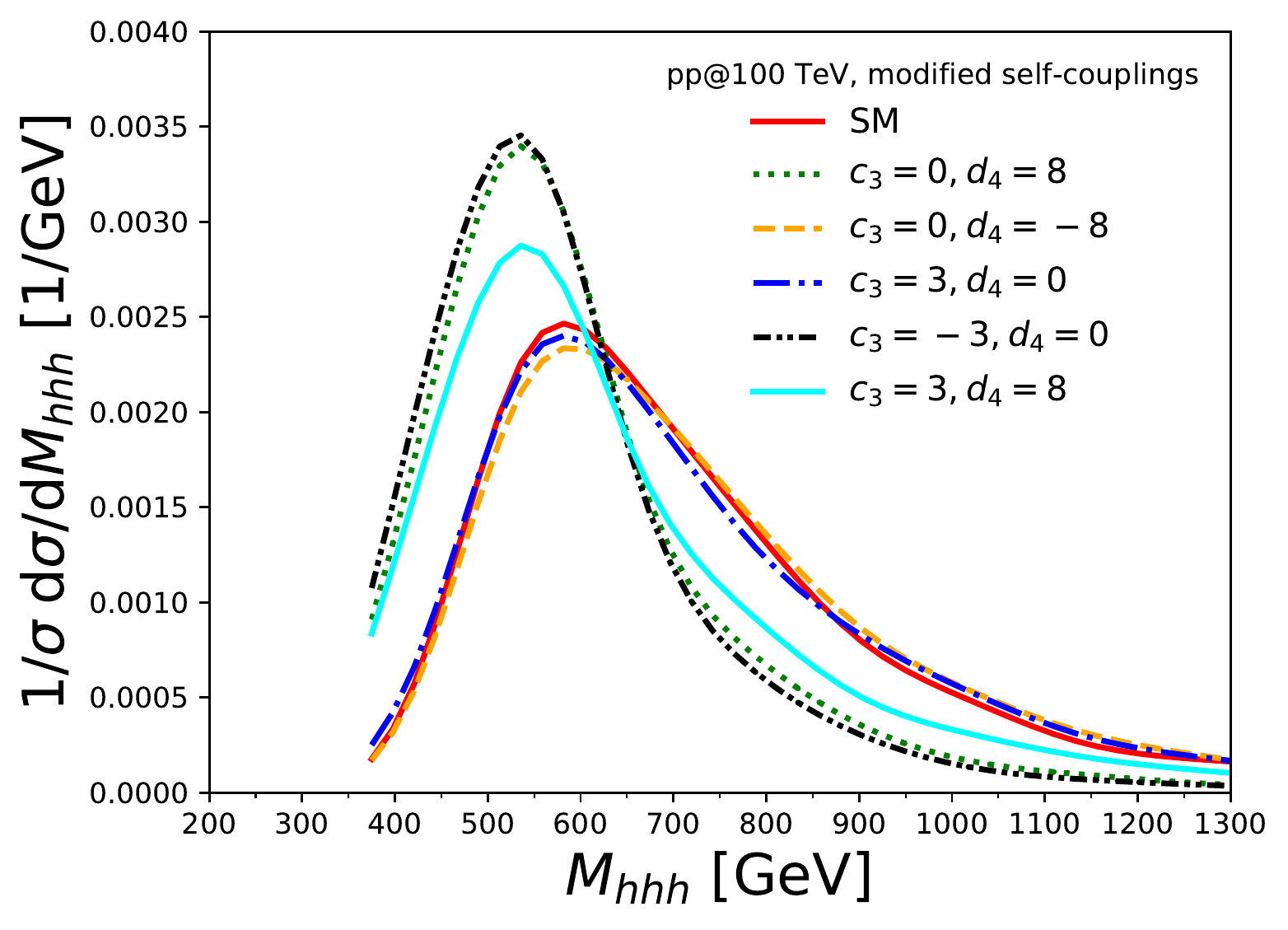}
\caption{The invariant mass of three Higgs bosons in triple Higgs production in the SM and with modified self-couplings ($\lambda_4 = \lambda_{\mathrm{SM}} (1+d_4)$ and $\lambda_3 = \lambda_{\mathrm{SM}} (1+c_3)$) at the FCC-hh at 100 TeV, reconstructed from Monte Carlo truth with no cuts applied.}
\label{fig:invmassd4orc3}
\end{figure}


We show in fig.~\ref{fig:invmassd4orc3} the normalised invariant mass distribution for the triple Higgs boson system for a few extreme values of the modifications of the quartic or triple coupling. It is clear that the anomalous couplings can modify substantially not only the cross section but also the distributions. This implies that for the same analysis cuts, the efficiency values will vary across the $(c_3, d_4)$-plane. The efficiency ranges from $\sim 0.5\%$ to $\sim 2\%$. A fit of the analysis efficiency $\epsilon (\geq 0)$, for the cuts given in table~\ref{tab:analysiscuts}, over the $(c_3, d_4)$-plane yields:
\begin{eqnarray}\label{eq:effc3d4}
\epsilon(c_3, d_4)  &=&  \left[0.01 \times c_3^4 - 0.01 \times c_3^3\right. \nonumber \\
&+&  0.00 \times c_3^2 d_4  - 0.10 \times c_3^2\nonumber\\
&+& 0.00 \times d_4^2 - 0.00 \times c_3 d_4\nonumber\\
&-& \left.0.02 \times d_4 + 0.20 \times c_3 + 1.31 \right] \% \;,
\end{eqnarray}
where some coefficients are zero up to the uncertainty obtained by the Monte Carlo sample employed. The cross section after cuts over the $(c_3, d_4)$-plane can be obtained by convolving eqs.~\ref{eq:sigmac3d4} and~\ref{eq:effc3d4}.

\begin{table*}[t!]
\caption{The processes considered in the six $b$-jet analysis, for the Standard Model. The second column shows the generation-level cross sections with the cuts (if any) as given in the main text. The $Z$ bosons were decayed at generation level and hence the cross section is given with the $Z$ branching ratios applied. The third column shows the starting cross section for the analysis, including the branching ratio to $(b\bar{b}) (b\bar{b}) (b\bar{b})$, with a flat $K$-factor of $K=2.0$ applied to all tree-level processes as an estimate of the expected increase in cross section from leading order to next-to-leading order. The fourth column gives the analysis efficiency and the final column gives the expected number of events at $20$~ab$^{-1}$ of integrated luminosity at 100 TeV. The results are given for perfect $b$-jet tagging efficiency. The label ``ggF'' implies that it is gluon-fusion initiated.}
\label{table:processes}
\begin{tabular*}{\textwidth}{@{\extracolsep{\fill}}lrrrl@{}}
\hline
Process& \multicolumn{1}{c}{$\sigma_\mathrm{GEN}$ (pb)} & \multicolumn{1}{c}{$\sigma_\mathrm{NLO}  \times \mathrm{BR} $ (pb)} & \multicolumn{1}{c}{$\epsilon_\mathrm{analysis}$ } & \multicolumn{1}{c}{$N^{\mathrm{cuts}}_{20~\mathrm{ab}^{-1}}$ }  \\
\hline
$hhh$ (SM)                                                                                  & $2.88 \times 10^{-3}$   & $1.06 \times 10^{-3}$ & 0.0131                           & 278 \\\hline
QCD $(b\bar{b}) (b\bar{b}) (b\bar{b})$                                         &    26.15                         & 52.30                          & $3.41 \times 10^{-5}$    & 35653 \\
$q\bar{q} \rightarrow hZZ\rightarrow h (b\bar{b}) (b\bar{b})$                                & $8.77 \times 10^{-4}$    & $4.99 \times 10^{-4}$ & $2.39 \times 10^{-4}$   & $\sim 2$ \\
$q\bar{q} \rightarrow ZZZ\rightarrow (b\bar{b}) (b\bar{b})$                                   &    $7.95 \times 10^{-4}$ & $7.95 \times 10^{-4}$ & $1.23 \times 10^{-5}$   & $\sim 1$ \\
ggF $hZZ\rightarrow h (b\bar{b}) (b\bar{b})$                               &   $1.08 \times 10^{-4}$  & $1.23 \times 10^{-4}$ & $\mathcal{O}(10^-3)$   & $\sim 2$ \\
ggF $ZZZ\rightarrow (b\bar{b}) (b\bar{b})$                                  & $1.36 \times 10^{-5}$    & $2.73 \times 10^{-5}$ & $\sim 4 \times 10^{-5}$ & $\ll 1$ \\
$h (b\bar{b}) (b\bar{b})$                                                               &   $1.46 \times 10^{-2}$                            &  $1.66 \times 10^{-2}$   & $5.4 \times 10^{-4}$    & 179 \\
$hh(b\bar{b})$                                                                              &  $1.40 \times 10^{-4}$  &  $9.11 \times 10^{-5}$ & $3.0 \times 10^{-5}$      & $\sim 1$ \\
$hhZ\rightarrow hh (b\bar{b})$                                                    &  $4.99 \times 10^{-3}$   & $1.61 \times 10^{-3}$ & $8.5 \times 10^{-4}$    & 27\\
$hZ(b\bar{b})\rightarrow h (b\bar{b}) (b\bar{b})$                          &  $9.08 \times 10^{-3}$   & $1.03 \times 10^{-2}$ & $1.5\times 10^{-4}$       & 31 \\
$ZZ(b\bar{b})\rightarrow (b\bar{b})(b\bar{b}) (b\bar{b})$              &  $2.87 \times 10^{-2}$   & $5.74 \times 10^{-2}$ & $1 \times 10^{-5}$      & 11 \\
$Z(b\bar{b})(b\bar{b})\rightarrow (b\bar{b})(b\bar{b}) (b\bar{b})$ &  0.93                               & $1.87$                       &  $1.4 \times 10^{-4}$      & 5233 \\\hline
$\sum$ backgrounds & \multicolumn{4}{r}{$2.8 \times 10^4$} \\
\hline
\end{tabular*}
\end{table*}

\subsection{Background processes}

Some of the background processes will be affected at the order we are considering by the rescaling of the self-couplings of eq.~\ref{eq:c3d4}, an effect that should be taken into account in a future analysis. However, we found that the processes that are affected at leading order by the anomalous couplings, i.e. those of the form $hh+X$, where $X=Z$ or $b\bar{b}$, constitute sub-permille contributions to the sum of all backgrounds after our analysis cuts are applied (see results of table~\ref{table:processes}). Therefore we do not consider these variations in our analysis, instead only considering their SM counterparts as an order-of-magnitude estimate. 

\begin{table}
\caption{The reducible background processes considered in the six $b$-jet analysis. The second column shows the generation-level cross sections with the cuts identical to the ones applied to the irreducible processes (table~\ref{tab:analysiscuts}). The third column shows the cross section after the mis-tagging rates have been applied. We only consider processes equivalent to QCD 6 $b$-jet production. We do not consider process that contain mis-tagged light and charm jets at the same time.}
\label{table:reducible}
\centering
\begin{tabular}{ccc}
process & $\sigma_\mathrm{GEN}$ (pb) & $\sigma_\mathrm{GEN}  \times \mathcal{P}(6~b-\mathrm{jets})$ (pb)\\
\noalign{\smallskip}\hline\noalign{\smallskip}
$(b\bar{b}) (b\bar{b}) (c\bar{c})$ & $76.8$ & $0.768$\\
$(b\bar{b}) (c\bar{c}) (c\bar{c})$ & $75.6$ & $0.00756$ \\
$(c\bar{c}) (c\bar{c}) (c\bar{c})$ & $22.5$ & $22.5\times 10^{-5}$ \\
$(b\bar{b}) (b\bar{b}) (jj)$ & $1.32 \times 10^4$ & $1.32$\\
$(b\bar{b}) (jj) (jj)$ & $9.79 \times 19^5$ & $0.00979$ \\
$(jj) (jj) (jj)$ & $1.37 \times 10^6$& $1.37 \times 10^{-6}$\\
\noalign{\smallskip}\hline
\end{tabular}
\end{table}

It is also evident that in table~\ref{table:processes} we have only included irreducible processes, those that are identical at parton level in flavour content to the signal: $(b\bar{b}) (b\bar{b}) (b\bar{b})$. As discussed previously, the degree of the contamination from reducible backgrounds, those that come from the mis-identification of light jets or charm-jets to $b$-jets, can be estimated by assuming that the efficiency of the analysis is identical to that of the equivalent irreducible ones. Explicitly, we will assume e.g. that the probability of a $(b\bar{b}) (b\bar{b}) (c\bar{c})$ event passing the analysis cuts is identical to $(b\bar{b}) (b\bar{b}) (b\bar{b})$, multiplied by the probability that two charm jets are mis-identified as $b$-jets. We will assume that the probability of a charm-jet being mis-identified as $b$-jet is $\mathcal{P}_{c\rightarrow b} = 0.1$ and that of light jets is $\mathcal{P}_{j\rightarrow b} = 0.01$, and that these values are independent of the $b$-tagging efficiency which we will take to range from perfect (100\%) to the ``worst-case scenario'' of 80\%, see~\ref{app:variations}.\footnote{We note that these rejection rates are close to those used in the self-coupling studies of Ref.~\cite{Benedikt:2018csr}. They are also not far from what is currently achievable with the ATLAS and CMS experiments, see e.g.~\cite{Chatrchyan:2012jua, Aad:2019aic}.} Table~\ref{table:reducible} shows the starting cross sections of the main reducible processes and the estimated contribution to the total cross section of the equivalent irreducible process, QCD six $b$-jet production by taking into account appropriate rescaling with powers of $\mathcal{P}_{c\rightarrow b}$ and $\mathcal{P}_{j\rightarrow b}$. Given our results, the reducible six-jet QCD backgrounds are expected to contribute $\mathcal{O}(10\%)$ to $\mathcal{O}(30\%)$ of the total tagged six $b$-jet background, for perfect $b$-tagging to $\mathcal{P}_{b\rightarrow b} = 0.8$, respectively. Therefore it is clear that the contributions are sub-dominant with respect to the irreducible process and from here on we absorb them in the overall uncertainty of the cross section estimates, the effect of which is also examined in~\ref{app:variations}. 

\subsection{Results for anomalous triple Higgs boson production}

\begin{figure}[t!]
  \includegraphics[width=\columnwidth]{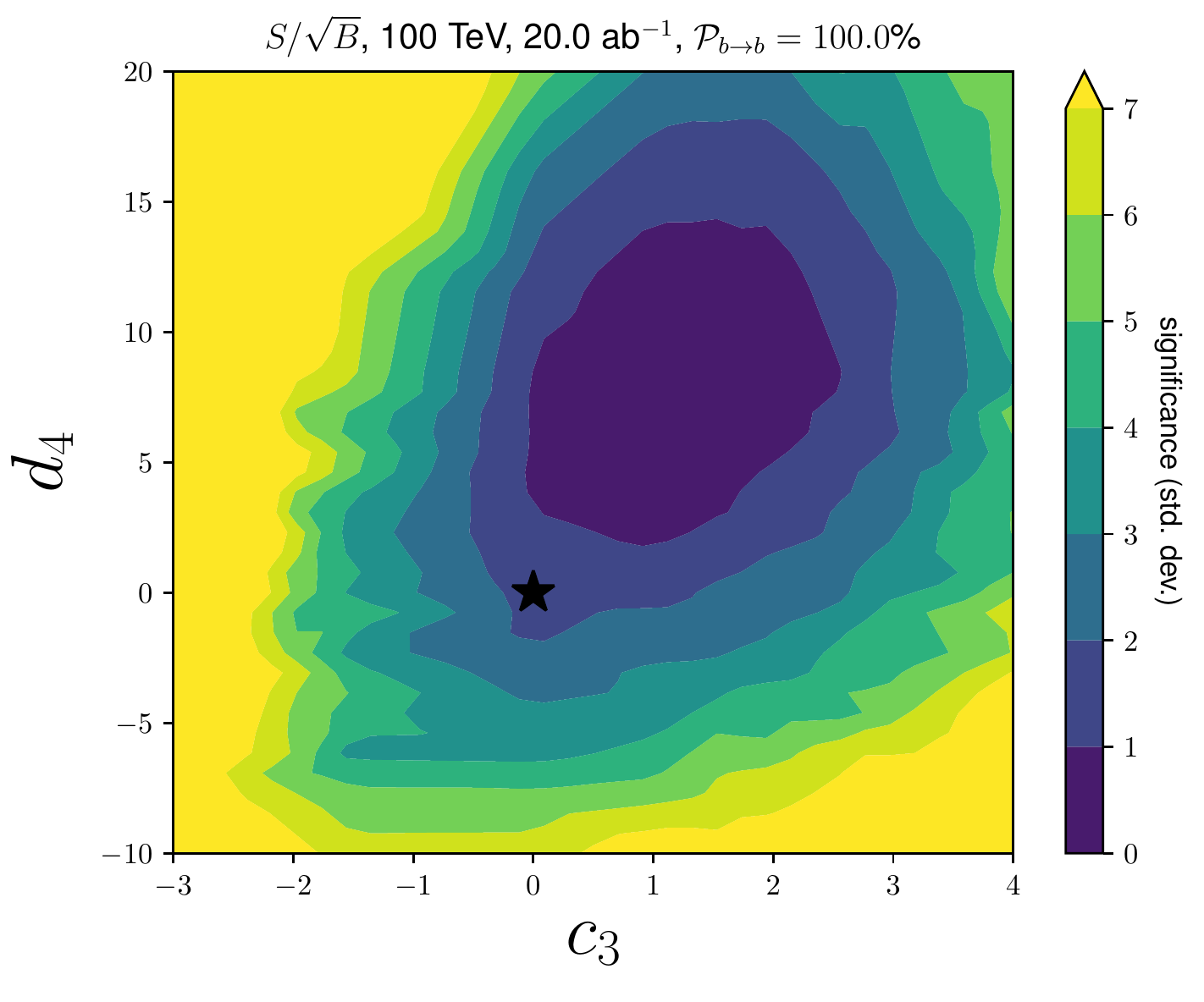}
\caption{The significance for our analysis of triple Higgs boson production in the $(b\bar{b}) (b\bar{b}) (b\bar{b})$ final state with modified self-couplings ($\lambda_4 = \lambda_{\mathrm{SM}} (1+d_4)$ and $\lambda_3 = \lambda_{\mathrm{SM}} (1+c_3)$) at the FCC-hh at 100 TeV. We have assumed an integrated luminosity of 20~ab$^{-1}$ and perfect $b$-tagging. The black star indicates the SM point.}
\label{fig:signifc3d4}
\end{figure}

\begin{figure}[t!]
  \includegraphics[width=0.9\columnwidth]{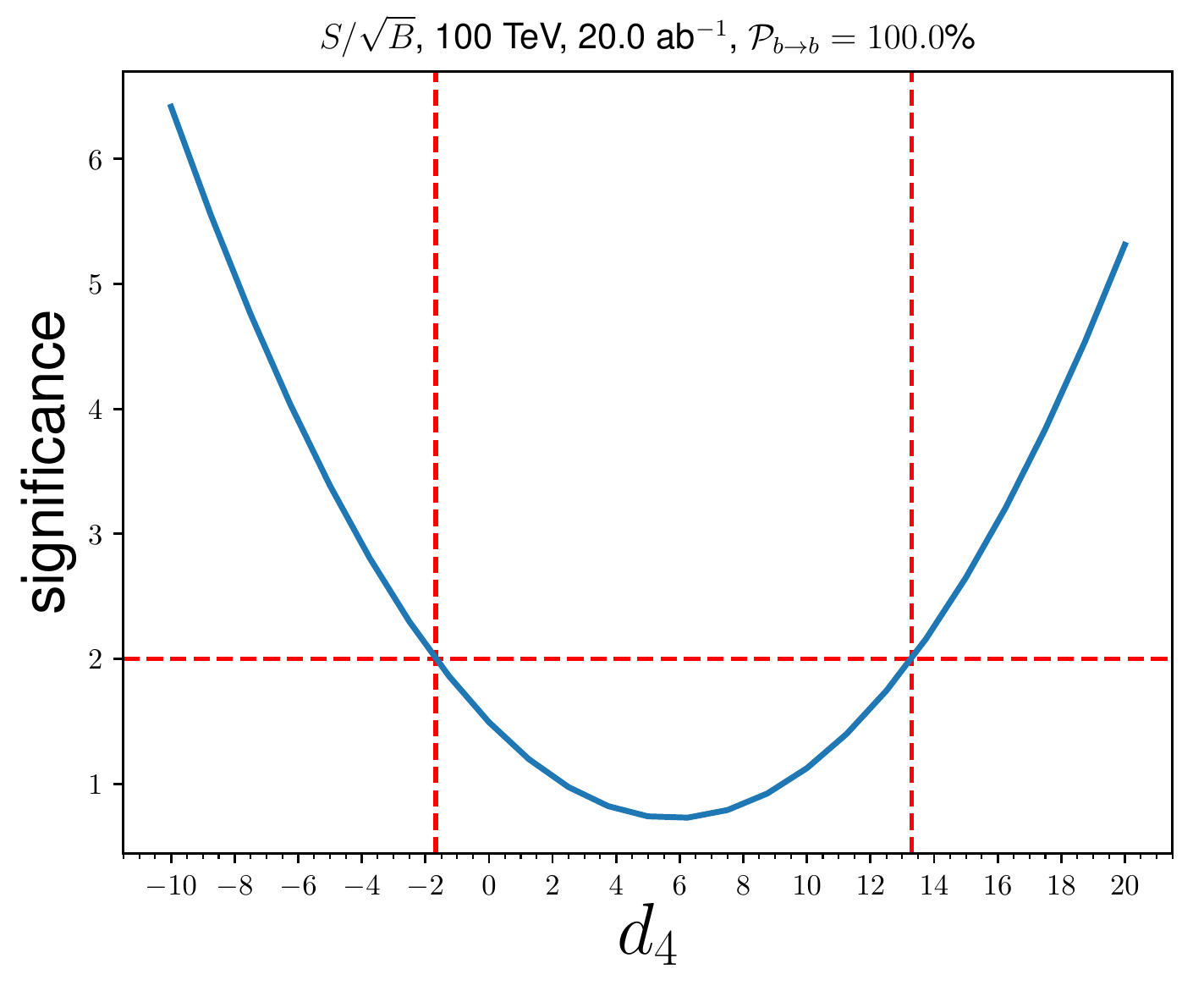}
\caption{The significance for our analysis of triple Higgs boson production in the $(b\bar{b}) (b\bar{b}) (b\bar{b})$ final state with modified quartic self-coupling ($\lambda_4 = \lambda_{\mathrm{SM}} (1+d_4)$) and no modification to the triple self-coupling ($c_3=0$) at the FCC-hh at 100 TeV. We have assumed an integrated luminosity of 20~ab$^{-1}$ and perfect $b$-tagging. The red dashed lines indicate the $2\sigma$ points for the constraints on $d_4$.}
\label{fig:signifd4only}
\end{figure}

As a result of the analysis described in subsection~\ref{sec:anal}, we show the expected significance that would be obtained on the $(d_4, c_3)$-plane for an integrated luminosity of 20~ab$^{-1}$ and assuming perfect $b$-tagging. Given that the constraints on the triple self-coupling at the FCC-hh will reach the percent level, we also consider a scenario in which $c_3=0$, allowing for variations of the quartic self-coupling through $d_4$. The resulting one-dimensional significance is shown in fig.~\ref{fig:signifd4only} for the case of perfect $b$-tagging. The constraint in this scenario would then be, at 95\% confidence level (i.e. $2\sigma$), $d_4 \in [-1.7, 13.3]$ as indicated by the red dashed lines in the figure. We defer the equivalent plots with reduced $b$-tagging efficiencies and the range of the pseudorapidity of $b$-tagging to~\ref{app:variations}. We note that the significance for the SM triple Higgs boson production ($d_4 = 0, c_3 = 0$) is $\sim 1.7\sigma$, up to the Monte Carlo uncertainties. 

\section{Triple Higgs boson production in the presence of a singlet scalar}\label{sec:singlet}

The discussion of the so-called xSM and the study of this section follows from~\cite{Kotwal:2016tex}. A more detailed discussion of the model and its relation to strong first-order electroweak phase transition is discussed therein. 

\subsection{The xSM}

The most general form of the xSM that depends on the Higgs doublet, $H$, and a gauge-singlet scalar, $S$, is given by (see, e.g.~\cite{OConnell:2006rsp, Profumo:2007wc, Barger:2007im, Espinosa:2011ax, Kotwal:2016tex}):
\begin{eqnarray}\label{eq:xsm}
V(H,S) = &-&\mu^2 (H^\dagger H) + \lambda (H^\dagger H)^2 + \frac{a_1}{2} (H^\dagger H) S \\ \nonumber
             &+& \frac{a_2}{2} (H^\dagger H) S^2 + \frac{b_2}{2} S^2 + \frac{b_3}{3} S^3 + \frac{b_4}{4} S^4 \;,
\end{eqnarray}
where the interactions proportional to $a_{1,2}$ constitute the Higgs ``portal'' that links the SM with the singlet scalar. We follow the study of~\cite{Kotwal:2016tex} in retaining all of the parameters, i.e. we do not impose a $\mathbb{Z}_2$ symmetry that would preclude terms of odd powers of $S$. 

After EWSB, the Higgs doublet and the singlet scalar both attain vevs: $H \rightarrow (v_0 + h) / \sqrt(2)$, with $v_0 \simeq 246$~GeV and $S \rightarrow x_0 + s$. Inevitably, the two states $h$ and $s$ mix through both the Higgs portal parameters $a_1$ and $a_2$ as well as the singlet vev. Diagonalising the mass matrix, one obtains two eigenstates, denoted by $h_1$ and $h_2$, where:
\begin{eqnarray}
h_1 &=& h \cos \theta + s \sin \theta \;, \\ \nonumber 
h_2 &=& - h \sin \theta + s \cos \theta \;. 
\end{eqnarray}
where $\theta$ is a mixing angle that can be expressed in terms of the parameters of the model. For $\theta \sim 0$, $h_1 \sim h$ and $h_2 \sim s$. We will identify the eigenstate $h_1$ with the state observed at the LHC, and hence set $m_1 = 125$~GeV. 


All the couplings of $h_{1,2}$ to the rest of the SM states are simply obtained by rescaling by:
\begin{equation}
g_{h_1 XX} = \cos \theta g_{hXX}^{\mathrm{SM}}\;,\;\; g_{h_2 XX} = \sin \theta g_{hXX}^{\mathrm{SM}}\;,
\end{equation}
with $XX$ any SM final state. This allows for constraints to be imposed on $\cos \theta$ through the measurements of Higgs signal strengths. We concentrate on the scenario $m_2 \geq 2 m_1$, allowing for resonant $h_2 \rightarrow h_1 h_1$, with no new decay modes appearing for the $h_1$. The triple couplings between the scalars $h_1$ and $h_2$, representing terms of the form $V(h_1, h_2) \supset \lambda_{ijk} h_i h_j h_k$, $i, j, k = \{1, 2\}$, are given by:
\begin{eqnarray}\label{eq:xsmtriple}
    \lambda_{111} &=& \lambda v_0 c_\theta^3 + \frac{1}{4} (a_1 + 2 a_2 x_0)  c_\theta^2 s_\theta\;,\\ \nonumber
                  &+& \frac{1}{2} a_2 v_0 s_\theta^2 c_\theta + \left(\frac{b_3}{3} + b_4  x_0\right) s_\theta^3 \;, \\ \nonumber
  \lambda_{112} &=& v_0 ( a_2 - 3 \lambda )  c_\theta^2 s_\theta - \frac{1}{2} a_2 v_0 s_\theta^3 \\ \nonumber
 &+& \frac{1}{2} ( -a_1 - 2 a_2 x_0 + 2 b_3 + 6 b_4 x_0 ) c_\theta s_\theta^2 + \frac{1}{4}  (a_1 + 2 a_2 x_0 ) c_\theta^3 \;, \\ \nonumber
    \lambda_{122} &=&  v_0  ( 3 \lambda - a_2 )  s_\theta^2  c_\theta + \frac{1}{2}  a_2 v_0 c_\theta^3 \\ \nonumber
&+& (b_3 + 3 b_4 x_0 - \frac{1}{2} a_1 - a_2 x_0 ) s_\theta c_\theta^2 + \frac{1}{4} ( a_1 + 2 a_2 x_0 )  s_\theta^3 \;, \\ \nonumber
    \lambda_{222} &=& \frac{1}{12}\left[4 (b_3 + 3 b_4 x_0)  c_\theta^3 - 6  a_2 v_0 c_\theta^2 s_\theta \right.\\ \nonumber
                  &+& \left.3 (a_1 + 2 a_2 x_0) c_\theta s_\theta^2 - 12 \lambda v_0 s_\theta^3 \right] \;,
\end{eqnarray}
where we have defined $c_\theta \equiv \cos\theta$ and $s_\theta \equiv \sin \theta$. The quartic couplings, representing terms of the form $V(h_1, h_2) \supset \lambda_{ijkl} h_i h_j h_k h_l$, $i, j, k, l = \{1, 2\}$, are given by:
\begin{eqnarray}\label{eq:xsmquartic}
  \lambda_{1111}  &= & \frac{1}{4} ( \lambda c_\theta^4 + a_2 c_\theta^2 s_\theta^2 + b_4 s_\theta^4) \;, \\ \nonumber
    \lambda_{1112} &=& -\frac{1}{2}  [-b_4+\lambda+(-a_2+b_4+\lambda) (2 c_\theta^2 - 1) ] c_\theta s_\theta \;,\\ \nonumber
    \lambda_{1122} &=& \frac{1}{16} \{a_2+3(b_4+\lambda) \\\nonumber 
                  &+& 3(a_2-b_4-\lambda) [(c_\theta^2-s_\theta^2)^2 - (s_\theta c_\theta)^2]\} \;, \\ \nonumber
    \lambda_{1222} &=& \frac{1}{4} [b_4-\lambda+(-a_2+b_4+\lambda) (c_\theta^2 - s_\theta^2)] s_\theta c_\theta \;, \\ \nonumber
    \lambda_{2222} &=& \frac{1}{4} (b_4 c_\theta^4 + a_2 c_\theta^2  s_\theta^2 + \lambda s_\theta^4) \;.
\end{eqnarray}
The above couplings will lead to processes with multiple $h_1$ and $h_2$ in the final state. 

In~\cite{Kotwal:2016tex}, the authors studied parameter-space points, satisfying conditions on the scalar sector of the xSM that lead to strong first-order electroweak phase transition (SFOEWPT). They then derived benchmark points taken from this allowed set that leads to enhanced resonant Higgs boson pair production, i.e. $h_2 \rightarrow h_1 h_1$, considering the phenomenological consequences, i.e. whether enhanced $h_1 h_1$ would be observed at future colliders, including a 100 TeV proton collider. One of the main conclusions was that such a collider could probe nearly all of the viable SFOWEPT-viable parameter space through this process, leading to a potential discovery of the xSM. 

Here we consider the benchmark points of~\cite{Kotwal:2016tex} in the context of (SM-like) triple Higgs boson production, $pp \rightarrow h_1 h_1 h_1$, which can potentially lead to a measurement of both the triple and quartic couplings in the xSM, in the event of discovery. Furthermore, there could be fine-tuned points in the xSM that lead to some of the scalar couplings being small. In that scenario, triple Higgs boson production could conceivably provide an alternative route for discovery of the xSM. We show in tables~\ref{table:benchmarksmax} and~\ref{table:benchmarksmin} in the next section the parameters for the benchmark points, which are labelled in~\cite{Kotwal:2016tex} as ``B1max'' to ``B11max'' and ``B1min'' to ``B11min''. 

\subsection{Triple Higgs boson production in the xSM}\label{sec:hhhxsm}

The process by which three $h_1$ scalars are produced via gluon fusion consists of diagrams identical to those that appear in fig.~\ref{fig:feyndiags}, with the addition that there exist diagrams with SM-like Higgs propagators (i.e. $h_1$ in this case) substituted by $h_2$. The strength of the interactions that appear in these diagrams is governed by the triple and quartic couplings of eqs.~\ref{eq:xsmtriple} and~\ref{eq:xsmquartic}. Note that the triple $h_1$ coupling, $\lambda_{111}$, will also be modified in the xSM. In general there will be an intricate interference pattern between all the contributing non-resonant and resonant diagrams. Our aim here is not to provide a detailed study of these effects; instead we investigate the observability of triple Higgs boson production, $pp \rightarrow h_1 h_1 h_1$, in the context of the six $b$-jet final state, focussing on the SFOEWPT benchmark points provided in~\cite{Kotwal:2016tex}, which appear in tables~\ref{table:benchmarksmax} and~\ref{table:benchmarksmin}.  For each point we also give the total triple $h_1$ production cross section as a ratio to the SM $hhh$, including the full (top or bottom quark) loop structure and interference effects. For comparison we have also calculated the total $h_1$ pair production cross section as a ratio to the SM $hh$. One can observe that the enhancement in $h_1 h_1 h_1$ production can be larger than the enhancement in $h_1 h_1$. 

\begin{table*}[t!]
  \begin{tabular}{cccccccccccccc}
    \hline
Benchmark & ~ $\cos \theta$ ~ & ~ $\sin \theta$ ~ & ~ $m_2$ ~ & ~ $\Gamma_{h_2}$ ~  & ~ $x_0$ ~ & ~ $\lambda$ ~ & ~ $a_1$ ~ & ~ $a_2$ ~ & ~ $b_3$ ~ & ~ $b_4$ ~ & ~ $\frac{\sigma(h_1 h_1 )}{\sigma(hh)_\mathrm{SM}}$ ~  & ~ $\frac{\sigma(h_1 h_1 h_1)}{\sigma(h h h)_\mathrm{SM}}$ ~  & \\          
           &               &               & (GeV) &     (GeV)        & (GeV) &           & (GeV) &       & (GeV) &       &      &  &  \\         
    \hline
B1max & 0.976 & 0.220 & 341 & 2.42 &  257 & 0.92 & -377 & 0.392 & -403 & 0.77 & 22.44 & 60.55  \\
B2max & 0.982 & 0.188 & 353 & 2.17 &  265 & 0.99 & -400 & 0.446 & -378 & 0.69 & 22.43 & 56.69 \\
B3max & 0.983 & 0.181 & 415 & 1.59 &  54.6 & 0.17 & -642 & 3.80 & -214 & 0.16 & 6.43 &  3.01 \\
B4max & 0.984 & 0.176 & 455 & 2.08 &  47.4 & 0.18 & -707 & 4.63 & -607 & 0.85 & 5.19 & 3.37 \\
B5max & 0.986 & 0.164 & 511 & 2.44 &  40.7 & 0.18 & -744 & 5.17 & -618 & 0.82 & 3.49 & 2.94 \\
B6max & 0.988 & 0.153 & 563 & 2.92 &  40.5 & 0.19 & -844 & 5.85 & -151 & 0.083 & 2.79 & 3.60 \\
B7max & 0.992 & 0.129 & 604 & 2.82 &  36.4 & 0.18 & -898 & 7.36 & -424 & 0.28 & 2.51 & 4.70\\
B8max & 0.994 & 0.113 & 662 & 2.97 &  32.9 & 0.17 & -976 & 8.98 & -542 & 0.53 & 2.28 & 4.91 \\
B9max & 0.993 & 0.115 & 714 & 3.27 &  29.2 & 0.18 & -941 & 8.28 & 497 & 0.38 & 1.98 & 2.68  \\
B10max & 0.996 & 0.094 & 767 & 2.83 &  24.5 & 0.17 & -920 & 9.87 & 575 & 0.41 & 1.95 & 2.35 \\
B11max & 0.994 & 0.105 & 840 & 4.03 & 21.7 & 0.19 & -988 & 9.22 & 356 & 0.83 & 1.76 & 1.03 \\
 \hline
  \end{tabular}
  \caption{\small Values of the various xSM independent and dependent parameters for each of the benchmark values chosen to maximize the $\sigma \cdot BR (h_2 \to h_1 h_1)$ at a 100~TeV proton collider, taken from~\cite{Kotwal:2016tex}. The ratio of cross sections of $h_1 h_1$ to SM $hh$ and of $h_1 h_1 h_1$ to $hhh$ production are given in the last two columns.}
\label{table:benchmarksmax}
\vspace{6mm}
\end{table*} 

\begin{table*}[t!]
  \begin{tabular}{cccccccccccccc}
    \hline
Benchmark & ~ $\cos \theta$ ~ & ~ $\sin \theta$ ~ & ~ $m_2$ ~ & ~ $\Gamma_{h_2}$ ~  & ~ $x_0$ ~ & ~ $\lambda$ ~ & ~ $a_1$ ~ & ~ $a_2$ ~ & ~ $b_3$ ~ & ~ $b_4$ ~ & ~ $\frac{\sigma(h_1 h_1 )}{\sigma(hh)_\mathrm{SM}}$ ~  & ~ $\frac{\sigma(h_1 h_1 h_1)}{\sigma(h h h)_\mathrm{SM}}$ ~  & \\         
           &               &               & (GeV) &     (GeV)        & (GeV) &           & (GeV) &       & (GeV) &       &       &   &  \\             
    \hline
B1min & 0.999 & 0.029 & 343 & 0.041 &  105 & 0.13 & -850 & 3.91 & -106 & 0.29 & 2.35 & 1.24 \\
B2min & 0.973 & 0.231 & 350 & 0.777 &  225 & 0.18 & -639 & 0.986 & -111 & 0.97 & 1.86 & 0.76 \\
B3min & 0.980 & 0.197 & 419 & 1.32 &  234 & 0.18 & -981 & 1.56 & 0.42 & 0.96 & 2.04 & 0.78\\
B4min & 0.999 & 0.026 & 463 & 0.0864 &  56.8 & 0.13 & -763 & 6.35 & 113 & 0.73 & 2.34 &  1.68  \\
B5min & 0.999 & 0.035 & 545 & 0.278 &  50.2 & 0.13 & -949 & 8.64 & 151 & 0.57 &  2.39& 2.86 \\
B6min & 0.999 & 0.043 & 563 & 0.459 &  33.0 & 0.13 & -716 & 9.25 & -448 & 0.96 & 2.42 & 3.90 \\
B7min & 0.984 & 0.180 & 609 & 4.03 &  34.2 & 0.22 & -822 & 4.53 & -183 & 0.57 & 1.72 & 0.75 \\
B8min & 0.987 & 0.161 & 676 & 4.47 &  30.3 & 0.22 & -931 & 5.96 & -680 & 0.43 & 1.64 & 0.75 \\
B9min & 0.990 & 0.138 & 729 & 4.22 &  27.3 & 0.21 & -909 & 6.15 & 603 & 0.93 & 1.68  & 0.91 \\
B10min & 0.995 & 0.104 & 792 & 3.36 &  22.2 & 0.18 & -936 & 9.47 & -848 & 0.66 & 1.81 & 1.31 \\
B11min & 0.994 & 0.105 & 841 & 3.95 &  21.2 & 0.19 & -955 & 8.69 & 684 & 0.53 & 1.76  & 0.94 \\
 \hline
  \end{tabular}
    \caption{\small Values of the various xSM independent and dependent parameters for each of the benchmark values chosen to minimize the $\sigma \cdot BR (h_2 \to h_1 h_1)$ at a 100~TeV proton collider, taken from~\cite{Kotwal:2016tex}. The ratio of cross sections of $h_1 h_1$ to SM $hh$ and of $h_1 h_1 h_1$ to $hhh$ production are given in the last two columns.}
\label{table:benchmarksmin}
\end{table*}

\begin{figure}
  \includegraphics[width=0.9\columnwidth]{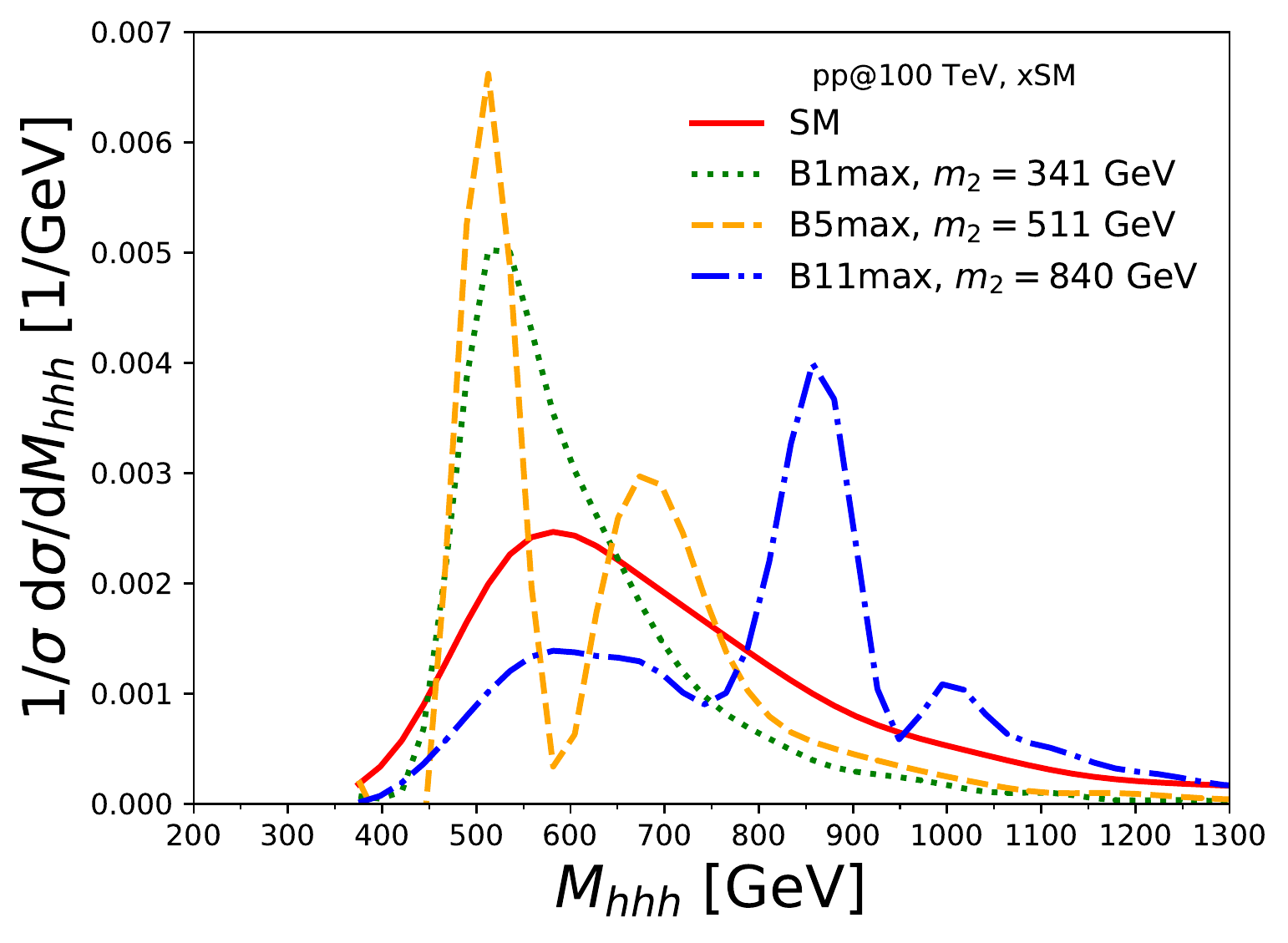}
\caption{The (normalised) invariant mass distribution of the three Higgs boson system in triple $h_1$ production within the xSM at the FCC-hh at 100 TeV, reconstructed from Monte Carlo truth with no cuts applied. We show three benchmark points from~\cite{Kotwal:2016tex}, as well as the SM expectation for comparison.}
\label{fig:invmassxsm}
\end{figure}

\begin{figure}
  \includegraphics[width=0.9\columnwidth]{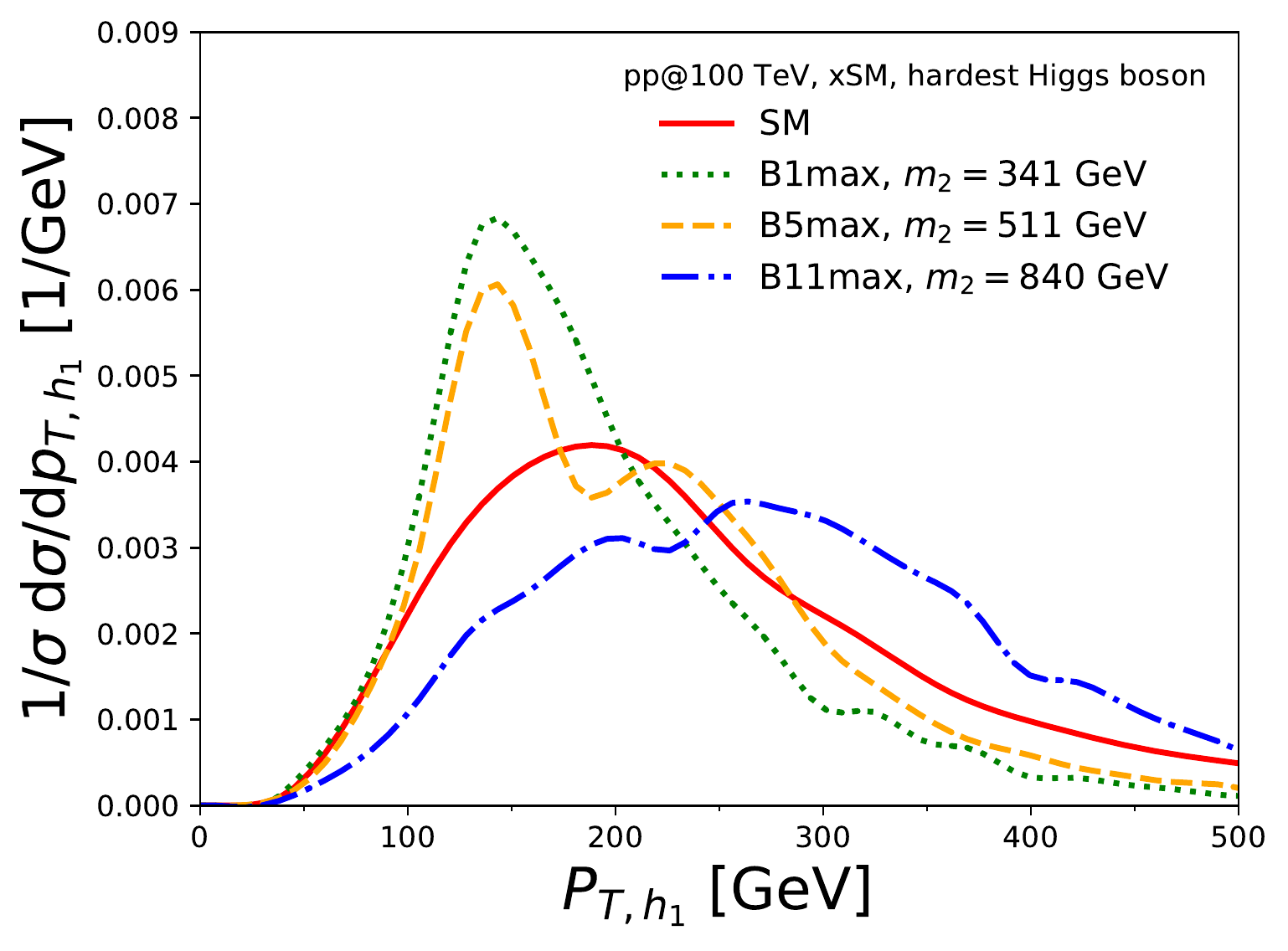}
\caption{The (normalised) transverse momentum distribution of the hardest Higgs boson in triple $h_1$ production within the xSM at the FCC-hh at 100 TeV, reconstructed from Monte Carlo truth with no cuts applied. We show three benchmark points from~\cite{Kotwal:2016tex}, as well as the SM expectation for comparison.}
\label{fig:pthiggsxsm}
\end{figure}

We show in figs.~\ref{fig:invmassxsm} and~\ref{fig:pthiggsxsm}, the invariant mass of the three Higgs boson system and the transverse momentum of the hardest Higgs boson in triple $h_1$ production within the xSM for three benchmark points as well as the SM expectation for comparison. The double-peak structure that is present in the distributions is physical and is due to the possibility of either an on-shell decay $h_2 \rightarrow h_1 h_1 h_1$, leading to a peak in $M_{hhh}$ at $\sim m_2$, or an on-shell decay $h_2 \rightarrow h_1 h_1$ with either $h_2$ or $h_1$ being off-shell in a preceding $s$-channel propagator, leading to the peak in $M_{hhh}$ at $\sim m_2 + m_1$. We note that a similar effect was pointed out in~\cite{Carmona:2016qgo} in $pp \rightarrow hS \rightarrow h \gamma\gamma$, in the context of a $\mathbb{Z}_2$-symmetric singlet scalar model.  

\subsection{Results for xSM triple Higgs boson production}

\begin{table}
\caption{The significance (in standard deviations) of the six $b$-jet analysis in $h_1 h_1 h_1$, applied to the benchmark points of tables~\ref{table:benchmarksmax} and~\ref{table:benchmarksmin}. We have assumed a perfect $b$-tagging efficiency.}
\label{table:xsmsignificance}
\centering
\begin{tabular}{cc|cc}
Benchmark & Significance & Benchmark  & Significance \\
\noalign{\smallskip}\hline\noalign{\smallskip}
B1max & 46.6 & B1min & 1.7\\
B2max & 42.9 & B2min& 1.3 \\
B3max & 2.9 & B3min & 1.1 \\
B4max & 3.7 & B4min & 2.0\\
B5max & 3.0 & B5min & 3.3 \\
B6max & 3.8 & B6min & 4.2 \\
B7max & 5.3 & B7min & 1.4 \\
B8max & 7.8 & B8min & 1.4 \\
B9max & 5.9 & B9min & 1.9 \\
B10max & 4.9 & B10min & 3.0 \\
B11max & 2.3 & B11min & 2.0 \\

\noalign{\smallskip}\hline
\end{tabular}
\end{table}

Table~\ref{table:xsmsignificance} shows the significance of the analysis applied to the 22 benchmark points B1max--B11max and B1min--B11min. The analysis has not been optimised for the specific features of these points, but the cuts are instead applied verbatim following those described previously (table~\ref{tab:analysiscuts}). It is quite likely, as was shown in~\cite{Kotwal:2016tex}, that the $h_2$ mass will be known during the lifetime of the FCC-hh through the observation of resonant production $ pp \rightarrow h_2 \rightarrow h_1 h_1$. This information could then be employed in the analysis to enhance the significance of the $h_1 h_1 h_1$ further, particularly taking into account the ``double-peak'' structure that we have pointed out in subsection~\ref{sec:hhhxsm}. Furthermore, cuts affected by the changes in the transverse momentum distributions as well as the angular distances can be subject to further optimisation. 

Given the values of the significance that we find here, it is conceivable that the $ pp \rightarrow h_1 h_1 h_1$ channel will play a crucial role in understanding the extended scalar sector in many viable scenarios of scalar gauge-singlet models that satisfy the constraints provided by requiring a SFOEWPT.\footnote{We note here that the ratio of $h_1 h_1 h_1$ to $h_1 h_1$ might be interesting to investigate in this scenario, so as to reduce theoretical uncertainties, as was done in~\cite{Goertz:2013kp} for the case of Higgs boson pair production.}

\section{Conclusions}\label{sec:conclusions}

We have investigated triple Higgs boson production at a future proton collider with centre-of-mass energy 100~TeV, in the case when all three Higgs bosons decay to bottom-anti-bottom quark pairs, producing six $b$-jets. We have constructed a detailed phenomenological hadron-level analysis including the effects of detector geometry. This analysis was applied to two scenarios: in the first, SM-like triple Higgs boson production, we allowed for ``anomalous'' modifications of the triple and quartic self-couplings independently. For the SM point, $(d_4, c_3) = (0,0)$, we demonstrated that significances of $\approx 2\sigma$ can be obtained from the six $b$-jet final state alone. Furthermore, we have shown that a constraint of $d_4 \gtrsim -2$ could be obtained in the case that the triple coupling is measured to be close to the SM value, $c_3 \sim 0$. These results are competitive with previously studied final states, rendering the six $b$-jet process an important contribution to the study of the self-couplings of the SM Higgs boson. In the second scenario, we considered an extension of the SM by a gauge-singlet scalar that could drive strong first-order electroweak phase transition. We investigated the triple production of the resulting SM-like scalar in the particular six $b$-jet final state, for several well-motivated benchmark points compatible with strong first-order electroweak phase transition, and we concluded that large significances can be obtained for many of these. This motivates further study of the triple Higgs boson process in the context of future collider studies of scalar singlet models. 

Finally, we emphasise that our conclusions are affected by uncertainties due to the absence of higher-order calculations for several of the background processes and details of the performance of the detector, particularly with respect to the tagging efficiencies, acceptance rates, resolution and triggers. Once these uncertainties have been better understood, a more detailed analysis, e.g. considering the differences between the radiation pattern of the colour singlet Higgs boson and QCD, or employing more advanced multivariate techniques, could lead to higher significances. Nevertheless, we have demonstrated here by varying several parameters, that the six $b$-jet process will certainly constitute an important component of the study of triple Higgs boson production at a future 100~TeV hadron collider. 

\begin{acknowledgements}
We would like to thank Olivier Mattelaer, Marieke Postma and Eleni Vryonidou for useful discussions. This work is supported by the Netherlands National Organisation for Scientific Research (NWO) that is funded by the Dutch Ministry of Education, Culture and Science (OCW). In particular, AP is supported by the NWO D-ITP consortium and GTX acknowledges support from the NWO program 156, ``Higgs as Probe and Portal''. AP additionally acknowledges support from the ERC grant ERC-STG-2015-677323.

\end{acknowledgements}

\appendix

\section{Variations and uncertainties}\label{app:variations}

\begin{figure}
  \includegraphics[width=\columnwidth]{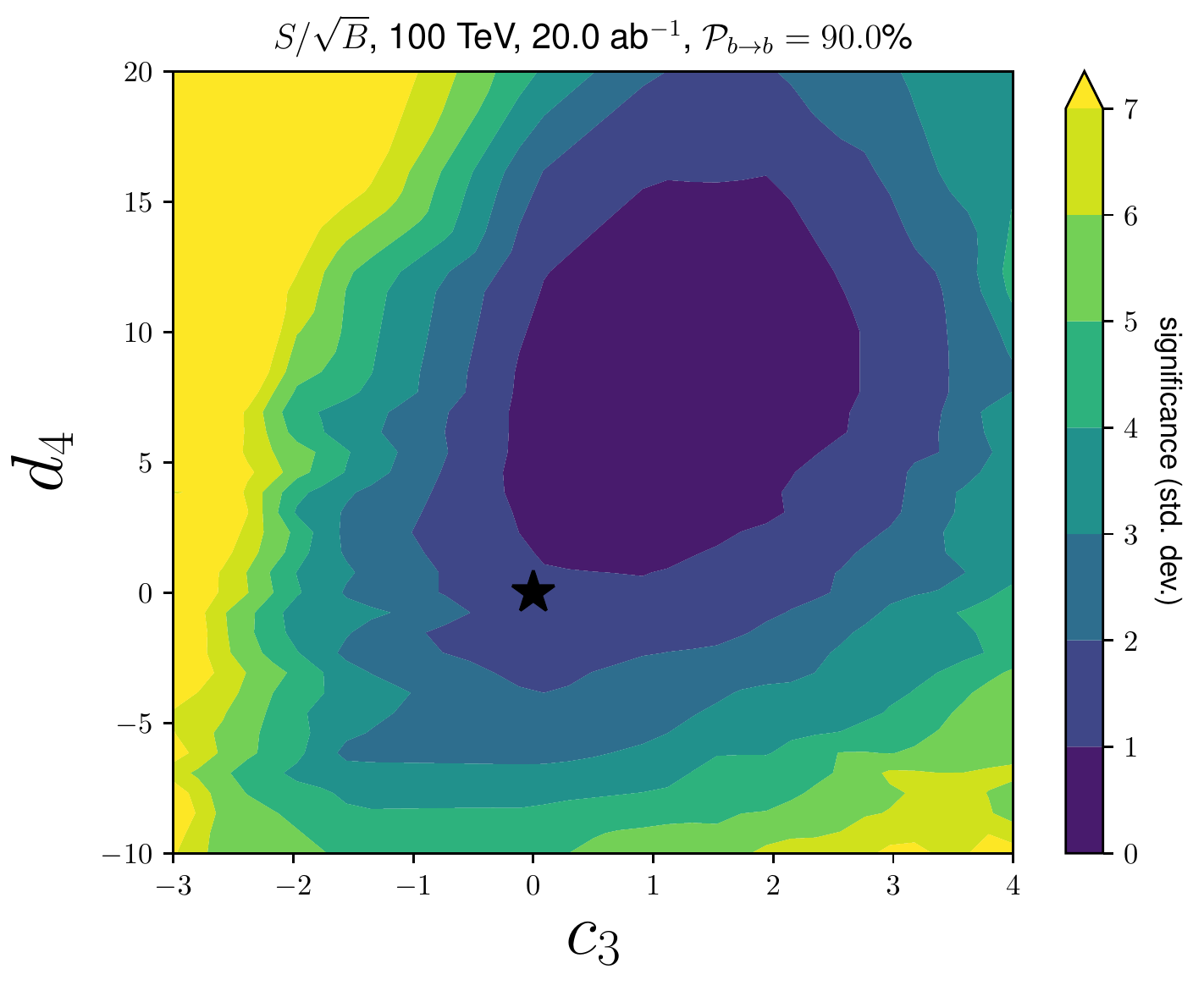}
\caption{The significance for our analysis of triple Higgs boson production in the $(b\bar{b}) (b\bar{b}) (b\bar{b})$ final state with modified self-couplings ($\lambda_4 = \lambda_{\mathrm{SM}} (1+d_4)$ and $\lambda_3 = \lambda_{\mathrm{SM}} (1+c_3)$) at the FCC-hh at 100 TeV. We have assumed an integrated luminosity of 20~ab$^{-1}$ and $b$-tagging efficiency of 90\%.  The black star indicates the SM point.}
\label{fig:signifc3d4_90perc}
\end{figure}

\begin{figure}
  \includegraphics[width=\columnwidth]{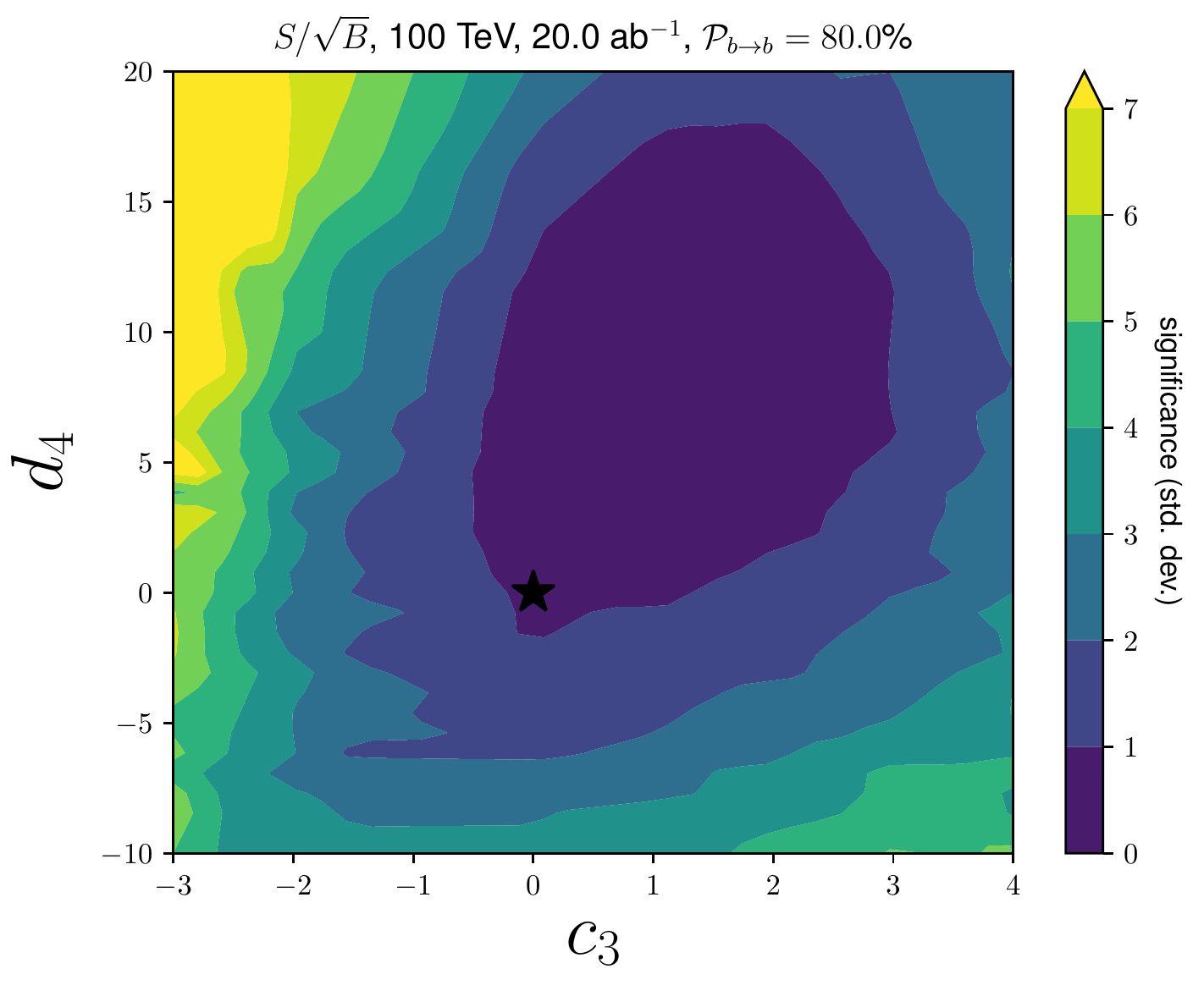}
\caption{The significance for our analysis of triple Higgs boson production in the $(b\bar{b}) (b\bar{b}) (b\bar{b})$ final state with modified self-couplings ($\lambda_4 = \lambda_{\mathrm{SM}} (1+d_4)$ and $\lambda_3 = \lambda_{\mathrm{SM}} (1+c_3)$) at the FCC-hh at 100 TeV. We have assumed an integrated luminosity of 20~ab$^{-1}$ and $b$-tagging efficiency of 80\%.  The black star indicates the SM point.}
\label{fig:signifc3d4_80perc}
\end{figure}

\begin{figure}
  \includegraphics[width=0.9\columnwidth]{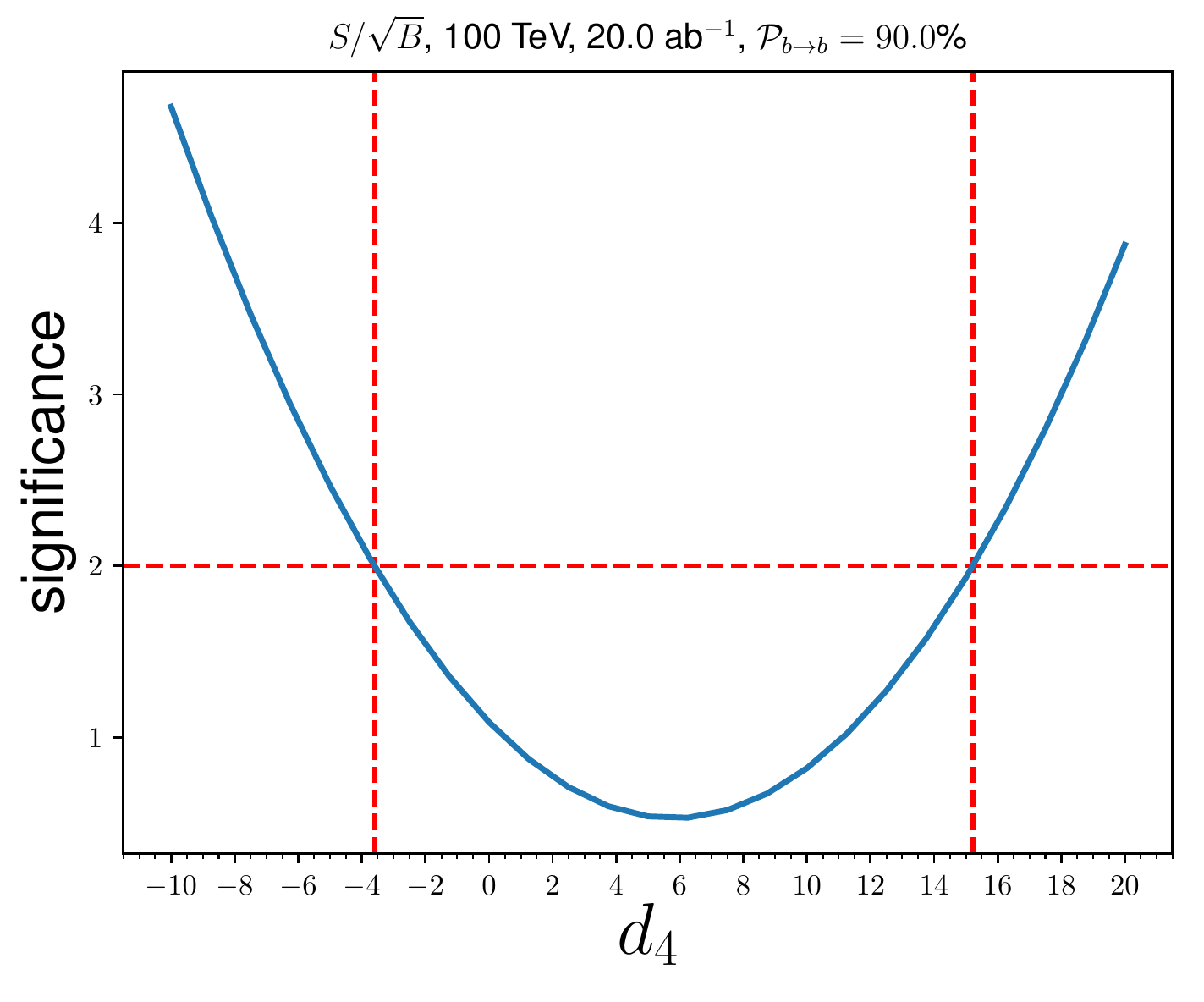}
\caption{The significance for our analysis of triple Higgs boson production in the $(b\bar{b}) (b\bar{b}) (b\bar{b})$ final state with modified quartic self-coupling ($\lambda_4 = \lambda_{\mathrm{SM}} (1+d_4)$) and no modification to the triple self-coupling ($c_3=0$) at the FCC-hh at 100 TeV. We have assumed an integrated luminosity of 20~ab$^{-1}$ and $b$-tagging of 90\%. The red dashed lines indicate the $2\sigma$ points for the constraints on $d_4$.}
\label{fig:signifd4only_90perc}
\end{figure}

\begin{figure}
  \includegraphics[width=0.9\columnwidth]{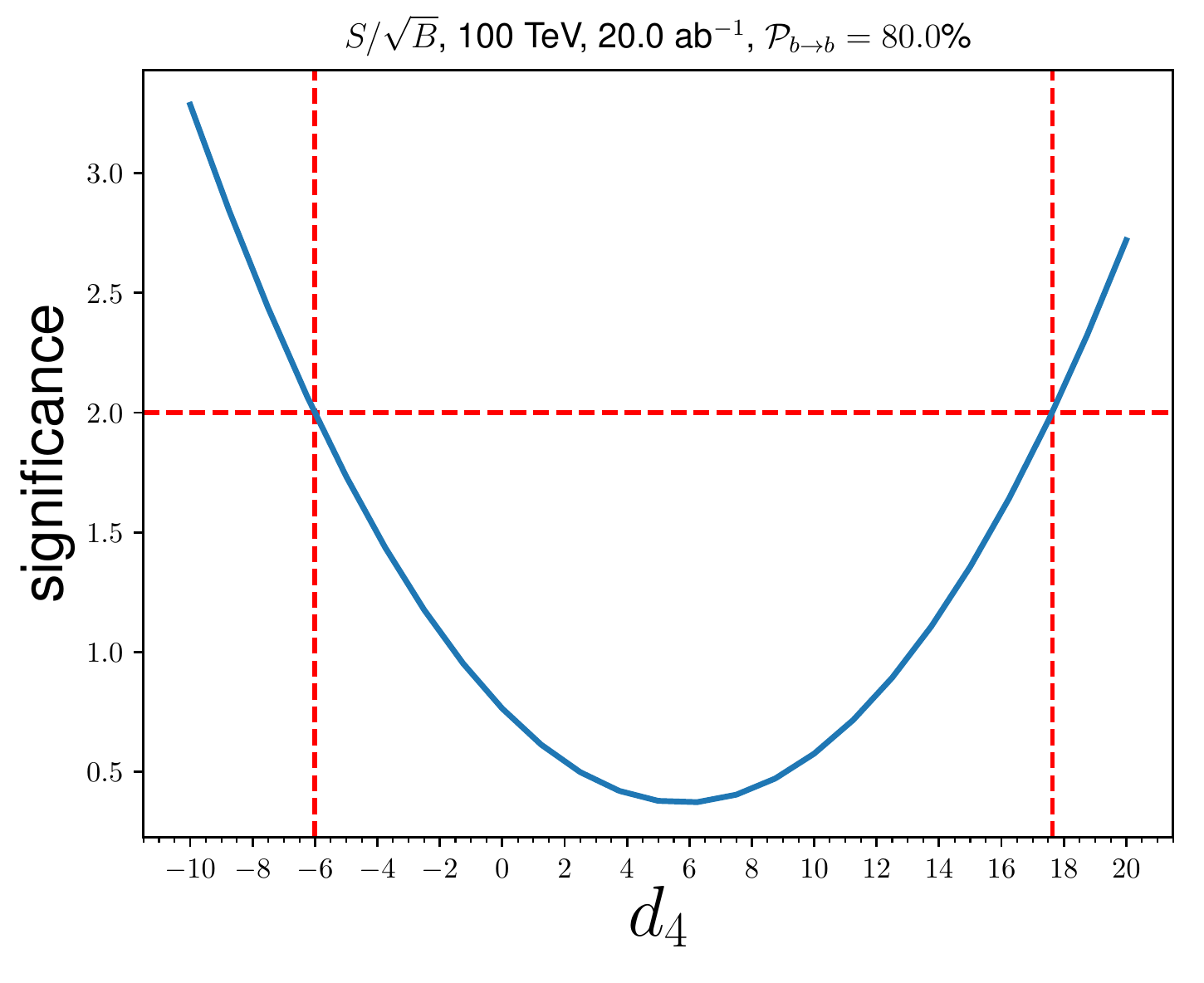}
\caption{The significance for our analysis of triple Higgs boson production in the $(b\bar{b}) (b\bar{b}) (b\bar{b})$ final state with modified quartic self-coupling ($\lambda_4 = \lambda_{\mathrm{SM}} (1+d_4)$) and no modification to the triple self-coupling ($c_3=0$) at the FCC-hh at 100 TeV. We have assumed an integrated luminosity of 20~ab$^{-1}$ and $b$-tagging of 80\%. The red dashed lines indicate the $2\sigma$ points for the constraints on $d_4$.}
\label{fig:signifd4only_80perc}
\end{figure}

In figs.~\ref{fig:signifc3d4_90perc} and~\ref{fig:signifc3d4_80perc} we show variations of the significance on the $(d_4, c_3)$-plane with 20~ab$^{-1}$, when the $b$-tagging efficiency is reduced from 100\% (perfect), to 90\% and 80\%, respectively. We also show the $c_3=0$ significance over the values of $d_4$ in figs.~\ref{fig:signifd4only_90perc} and~\ref{fig:signifd4only_80perc}. The equivalent constraints at 95\% C.L. on $d_4$ would then be, respectively, $d_4 \in [-3.6, 15.2]$ and $d_4 \in [-6.0, 17.6]$, with 20~ab$^{-1}$. 



The FCC-hh detector coverage over which $b$-jets will be tagged might also be tighter. Maintaining perfect $b$-tagging within the restricted region, with the given set of cuts, we apply $|\eta_b| < 2.5$ instead of $|\eta_b| < 3.2$. The 95\% C.L. constraint on $d_4$ in this case would be slightly shifted, yielding a range:  $d_4 \in [-3.1, 14.2]$ at 95\% C.L.. The significance for the SM point $(d_4, c_3) = 0$ is also reduced to $1.3\sigma$. 

Finally, by far the largest theoretical uncertainty is in the $K$-factors for the tree-level background processes. In the main part of this article, we have applied $K=2$ to all of these. If this is increased to $K=3$ for all tree-level background processes, we would obtain $\sim 4.3 \times 10^4$ background events, yielding a significance for the SM point of $\sim 1.3 \sigma$. On the other hand, if this is reduced $K=1.5$, the number of background events would decrease to $\sim 2.1 \times 10^4$ with a significance of $\sim 1.9\sigma$. We reckon that shape uncertainties may also be important, in particular those 
effects due to extra radiation generated \emph{before} the production of $b$ quarks. However given the complexity related to assessing this kind of uncertainties,
we do not include them in this explorative study.

\bibliography{hhh6b}
\bibliographystyle{JHEP}

\end{document}